\documentclass[reprint,amsmath,amssymb,aps,prapplied,floatfix]{revtex4-2}

\usepackage{lineno}
\usepackage{graphicx}
\usepackage{dcolumn}
\usepackage{bm}
\usepackage[T1]{fontenc}
\usepackage{xcolor}
\usepackage[none]{hyphenat}
\usepackage{tikz}
\usetikzlibrary{positioning,shapes.geometric,arrows.meta,matrix}
\usepackage{array}
\newcolumntype{C}[1]{%
    >{\centering\arraybackslash}m{#1}%
}
\usepackage{makecell}
\usepackage{multirow}
\begin{document}

\preprint{APS/123-QED}

\title{Parasitic Interference in Heterodyne Interferometers: Modeling, Characterization, and Mitigation}

\author{Pengzhuo Wang}
\author{Jose Sanjuan}
\author{Moritz Mehmet}
 \altaffiliation{Present address: Leibniz University Hannover, 30167 Hannover, Germany and Max Planck Institute for Gravitational Physics (Albert Einstein Institute), 30167 Hannover, Germany}
\author{Felipe Guzman}
 \email{felipeguzman@arizona.edu}
\affiliation{James C. Wyant College of Optical Sciences, The University of Arizona, 1630 E. University Blvd., Tucson, AZ 85721, USA}

\begin{abstract} 
Parasitic interference is a common limitation in laser interferometers, arising from unwanted beams that corrupt the phase measurement and degrade displacement sensitivity. In this work, we present a unified framework for the characterization and mitigation of parasitic interference in heterodyne interferometers. Parasitic beams are classified into two types based on the orientation of their corresponding phase vectors, and their noise contribution is modeled as a function of polarization, relative amplitude, and phase of the parasitic beam. Central to this framework is the concept of coupling coefficients, which quantify the interferometer's susceptibility to parasitic interference and can be readily computed using Jones calculus for any optical configuration. The two types of parasitic interference motivate distinct mitigation strategies: differential interferometry and balanced detection, complemented by polarization control and high-quality beam splitters. The models and mitigation strategies are validated experimentally in both a simplified Mach-Zehnder interferometer and a differential interferometer used for optomechanical inertial sensing, demonstrating reductions in parasitic phase noise by up to four orders of magnitude and achieving sub-picometer displacement sensitivity at frequencies as low as 9\,mHz under ambient atmospheric pressure. 
\end{abstract}

\maketitle
\section{Introduction} \label{S1}
Laser interferometers provide high-precision relative displacement measurements and are used across a wide range of applications, from gravitational wave detectors \cite{abbott2009ligo, acernese2015advanced, Akutsu2019Kagra,amaro2017laser} to space geodesy~\cite{abich2019orbit}, semiconductor lithography~\cite{mi16010006}, and large-scale optical instruments such as telescope segment alignment~\cite{photonics12121181}, to name a few. For our particular application, laser interferometers provide the displacement measurement in optomechanical inertial sensors~\cite{hines2020optomechanical,hines2023compact}, where they must resolve sub-picometer motion of a resonant test mass over a dynamic range of hundreds of micrometers, with stringent requirements on linearity and noise performance.

Interferometer implementations vary as widely as their applications, both in their optical layout and in the modulation scheme used to retrieve the interferometric signal. Classic optical configurations include the Michelson, Mach-Zehnder, Sagnac, and Fabry-Perot interferometers, along with many variants. In terms of modulation scheme, interferometers can be broadly divided into homodyne and heterodyne: in homodyne systems, the interfering beams share the same frequency, while in heterodyne systems, they are separated by a frequency offset known as the heterodyne frequency. Regardless of the configuration, the signal of interest is the relative phase between the two beams traveling different optical paths, which can be converted to displacement using the optical wavelength. In practice, however, any spurious beams that interfere with the main beams corrupt the phase measurement and can limit the sensitivity of the interferometer, a challenge common to all configurations.

Parasitic interference arises from unwanted beams generated by, e.g., multiple reflections at optical surfaces, optical birefringence, or radiofrequency (RF) crosstalk between modulators that interfere with the main beams or with each other. In heterodyne systems, the most problematic are those that beat at the heterodyne frequency, as they generate spurious signals that corrupt the phase measurement and can ultimately limit the sensitivity of the interferometer, a phenomenon known in the literature as periodic nonlinearity, cyclic error, or small vector noise~\cite{NBobroff_1993,Wu:99,armano2022sensor}. What makes parasitic interference particularly challenging is that the spurious beams are typically weak and difficult to isolate, their phase drifts with environmental conditions such as temperature and vibration, and their coupling to the main phase measurement is nonlinear, making systematic characterization and mitigation non-trivial.

The problem of parasitic interference in heterodyne interferometers has been studied previously, though typically for specific configurations or applications. Early work established the connection between ghost beams and periodic nonlinearity~\cite{NBobroff_1993,Wu:99,WU200317,SCHMITZ2003311,DeFreitas01091995,Keem:04}, and subsequent studies developed more detailed models of ghost reflection and its coupling with optical mixing~\cite{s18030758,WU200317,DeFreitas01091995}. Studies focused on LISA and LISA Pathfinder have characterized the effect of backreflections in the telescope~\cite{Spector_2012,Livas_2017,Sasso_2019}, optical bench~\cite{PhysRevLett.122.081104}, and fiber links~\cite{Isleif_2017,Isleif_2018,Fleddermann_2018} for those specific configurations. Scattered light has also been identified as a dominant noise source in ground-based gravitational wave detectors, with significant efforts devoted to its characterization and mitigation in Advanced LIGO and VIRGO~\cite{soni2021reducing,soni2024modeling,Was_2021,Longo_2024}. Mitigation strategies have also been explored more broadly, including real-time compensation algorithms~\cite{SCHMITZ2009353,GUO2022110334}, ghost beam suppression in deep frequency modulation interferometry~\cite{s21051708}, and tunable coherence techniques achieving up to 40\,dB of stray light suppression~\cite{PhysRevLett.134.213802}. 

In this work, we build on these efforts by providing a unified framework for the characterization and mitigation of parasitic interference in heterodyne interferometers. Parasitic interference is classified into type-I and type-II according to the relative orientations of the corresponding phase vectors, resulting in distinct noise behaviors. For type-I interference, each parasitic phase vector maintains the same angular separation from its corresponding main phase vector across all detection channels, whereas the type-II parasitic phase vectors are mutually parallel. A rigorous definition based on the phase-vector representation is provided in Sec.~\ref{S2}. Their noise contribution is modeled as a function of polarization, relative amplitude, and phase. Central to this framework is the concept of coupling coefficients, which quantify the interferometer's susceptibility to parasitic interference and can be readily computed using Jones calculus for any optical configuration. The two types naturally motivate distinct mitigation strategies, offering a path toward systematic elimination of parasitic coupling.

We validate the models and mitigation strategies experimentally in both a simplified Mach-Zehnder interferometer and a differential interferometer typically used for optomechanical inertial sensing. In the latter, parasitic phase noise is reduced by up to four orders of magnitude, enabling sub-picometer displacement sensitivities at frequencies above 9 mHz and phasemeter-limited noise levels of $\rm 30\,fm/\sqrt{Hz}$ between 2\,Hz and 10\,Hz, operating entirely under ambient atmospheric pressure.

The paper is organized as follows: Sec.~\ref{S2} describes the effect of parasitic beams on the interferometer readout depending on their origin and frequency, defining type-I and type-II parasitic interference. This section includes experimental validation of the models using simplified interferometer configurations to isolate the effects under investigation. In Sec.~\ref{S3}, the models and mitigation strategies developed in Sec.~\ref{S2} are applied to a differential interferometer used as the displacement readout of an optomechanical inertial sensor~\cite{hines2023compact}. Sec.~\ref{frame work} summarizes the mitigation strategies and presents a systematic framework for future interferometer design. We close with a summary of our findings in Sec.~\ref{S4}.

\section{Parasitic beams and interference} \label{S2}
In a heterodyne interferometer, two main beams separated by the heterodyne frequency ($\Delta \Omega = \Omega_1 - \Omega_2=2\pi\Delta f$) enter the optical system through two inputs, travel along two distinct optical paths, and accumulate phase shifts accordingly. The two beams recombine at the beam-splitter (BS), generating a beat-note at the heterodyne frequency, which is detected at the photodetectors (PD). Parasitic interference occurs when additional beam pairs beat at the heterodyne frequency, generating spurious signals that degrade the main phase measurement. Parasitic beams arise from various sources depending on the setup, e.g., stray light, ghost beams, acousto-optical modulators (AOM), radio-frequency (RF) crosstalk, and optical birefringence, to name a few. These beams experience additional phase shifts and amplitude fluctuations throughout the system, which couple nonlinearly into the retrieved phase signal.
\begin{figure}[ht] 
\centering\includegraphics[width = .9\linewidth]{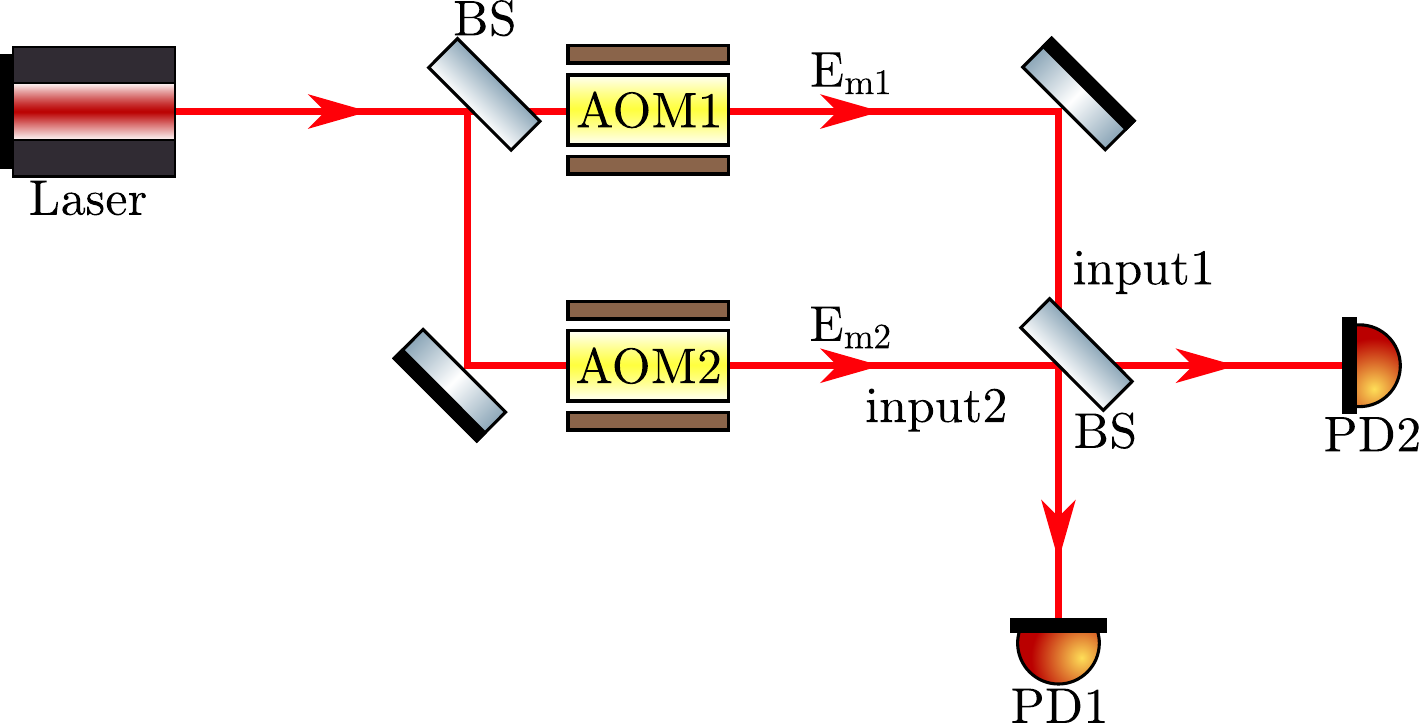}
\caption{Heterodyne Mach-Zehnder interferometer. BS: beam splitter. AOM: acousto-optic modulator. PD: photodetector.} \label{IFO MZ}
\end{figure}

To model the effect of parasitic beams, let us consider the heterodyne Mach-Zehnder interferometer shown in Fig.~\ref{IFO MZ}, where two AOMs are used to shift the frequency of each beam such that they differ by the heterodyne frequency $\Delta \Omega$. The two main beams ($\vec{E}_{\rm m_1}, \vec{E}_{\rm m_2}$) recombine at the BS, and the relative phase is measured at the two output ports on PD1 and PD2, respectively. The so-called $\pi$ test consists of taking the difference between the phase measured at PD1 and PD2, which ideally equals exactly $\pi$. However, the presence of parasitic beams ($\vec{E}_\epsilon$) at either frequency of the main beams prevents the result from being $\pi$. Thus, the $\pi$ test serves as a useful diagnostic for evaluating parasitic interference. 

Consider a parasitic beam entering the recombination BS from input 1 at the same frequency as the main beam 1 ($\Omega_1$). The electric field of each beam, characterized by its phase ($\phi$), amplitude ($|\vec{E}|$), and polarization direction ($\hat{\theta}$), can be expressed as
\begin{align}
 \vec{E}_{\rm m_1} &= |\vec{E}_{\rm m_1}|\hat{\theta}_{\rm m_1}e^{i[(\omega+\Omega_1)t+\phi_{\rm m_1}]}\\
 \vec{E}_{\rm m_2} &= |\vec{E}_{\rm m_2}|\hat{\theta}_{\rm m_2}e^{i[(\omega+\Omega_2)t+\phi_{\rm m_2}]}\\
 \vec{E}_{\epsilon} &= |\vec{E}_{\epsilon}|\hat{\theta}_{\epsilon}e^{i[(\omega+\Omega_1)t+\phi_{\rm m_1}+\phi_{\epsilon}]},
\end{align}
where $\hat{\theta}$ is the unit vector defining the angle between the polarization direction of the input beam and the s-polarization axis, where the p- and s-polarization directions are defined with respect to the plane of incidence at the recombination BS.
Without loss of generality, we assume the parasitic beam enters the BS through input 1 and, for convenience, define its phase as $\phi_{\rm m_1}+\phi_{\epsilon}$. We further assume both main beams share the same polarization and amplitude, i.e., $\hat{\theta}_{\rm m_1}=\hat{\theta}_{\rm m_2}=\hat{\theta}_{\rm m}$ and $|\vec{E}_{\rm m_1}|=|\vec{E}_{\rm m_2}|=|\vec{E}_{\rm m}|$. The electric fields at the inputs of the recombination BS are
\begin{align} \label{eq8}
 \vec{E}_{1} &= \vec{E}_{\rm m_1}+\vec{E}_{\epsilon}\\
 \vec{E}_{2} &= \vec{E}_{\rm m_2}.
\end{align}

The recombination BS plays an important role in parasitic interference coupling and overall interferometer performance. Beamsplitters exhibit polarization dependence such that the splitting ratio and losses differ for p- and s-polarized beams. Unless otherwise stated, we assume the following values of power reflectance and transmittance for our calculations throughout this section
\begin{align*}
 T_{1s} &= 33.01\%~~~~~T_{2s} = 31.43\%\\
 T_{1p} &= 63.61\%~~~~~T_{2p} = 63.38\%\\
 R_{1s} &= 62.93\%~~~~~R_{2s} = 62.20\%\\
 R_{1p} &= 34.94\%~~~~~R_{2p} = 32.26\%,
\end{align*} 
which have been measured from a commercial off-the-shelf BS. The corresponding transmission ($t$) and reflection ($r$) coefficients are the square roots of the values above. Thus, the electric fields at the PDs are
\begin{align}
 \vec{E}_{\rm PD_1} &= \vec{E}_{1}\cdot(t_{1s}\hat{s}+t_{1p}\hat{p})+\vec{E}_2\cdot(r_{2s}\hat{s}+r_{2p}\hat{p})e^{i\pi/2}\\
 \vec{E}_{\rm PD_2} &= \vec{E}_{1}\cdot(r_{1s}\hat{s}+r_{1p}\hat{p}) e^{i\pi/2}+\vec{E}_2\cdot(t_{2s}\hat{s}+t_{2p}\hat{p}),
\end{align}
where the $\pi/2$ phase shifts are introduced by the reflection at the BS. After algebraic manipulation of the preceding equations, calculating the power at the photodetector, and removing the DC terms, one obtains:
\begin{eqnarray}
S_1&=&|\vec{E}_{\rm m}|^2(\cos\theta_{\rm m}t_{1s}\cos\theta_{\rm m}r_{2s}+\sin\theta_{\rm m}t_{1p}\sin\theta_{\rm m}r_{2p})\nonumber\\
&&\times\cos(\Delta\Omega t+\Delta\phi_{\rm m}-\pi/2) \nonumber \\
 &&+|\vec{E}_{\epsilon}||\vec{E}_{\rm m}|(\cos\theta_{\epsilon}t_{1s}\cos\theta_{\rm m}r_{2s}+\sin\theta_{\epsilon}t_{1p}\sin\theta_{\rm m}r_{2p})\nonumber\\
 &&\times\cos(\Delta\Omega t+\Delta\phi_{\rm m}+\phi_{\epsilon}-\pi/2) \label{eq.8}\\
S_2&=&|\vec{E}_{\rm m}|^2(\cos\theta_{\rm m}r_{1s}\cos\theta_{\rm m}t_{2s}+\sin\theta_{\rm m}r_{1p}\sin\theta_{\rm m}t_{2p})\nonumber\\
&&\times\cos(\Delta\Omega t+\Delta\phi_{\rm m}+\pi/2) \nonumber \\
 &&+|\vec{E}_{\epsilon}||\vec{E}_{\rm m}|(\cos\theta_{\epsilon}r_{1s}\cos\theta_{\rm m}t_{2s}+\sin\theta_{\epsilon}r_{1p}\sin\theta_{\rm m}t_{2p})\nonumber\\
 &&\times\cos(\Delta\Omega t+\Delta\phi_{\rm m}+\phi_{\epsilon}+\pi/2), \label{eq.9}
\end{eqnarray}
where $\Delta\phi_{\rm m} = \phi_{\rm m_1}-\phi_{\rm m_2}$, and $\theta$ is the angle between the polarization of the beams and the s-polarization axis. The signals at $\Delta \Omega$ in Eqs.~(\ref{eq.8}) and (\ref{eq.9}) can be represented as vectors in the complex plane. Normalizing by their amplitude $|\vec{E}_{\rm m}|^2$, one obtains
\begin{eqnarray}
\vec{V}^{\rm I}_1&=&\vec{V}_{\rm m_1}+\vec{V}^{\rm I}_{\epsilon_1}\nonumber\\
&=&e^{i(\Delta\phi_{\rm m}-\pi/2)}+f_1^{\rm I}(\theta)\frac{|\vec{E}_{\epsilon}|}{|\vec{E}_{\rm m}|}e^{i(\Delta\phi_{\rm m}+\phi_{\epsilon}-\pi/2)} \label{eq.10} \\
\vec{V}^{\rm I}_2&=&\vec{V}_{\rm m_2}+\vec{V}^{\rm I}_{\epsilon_2}\nonumber\\
&=&e^{i(\Delta\phi_{\rm m}+\pi/2)}+f_2^{\rm I}(\theta)\frac{|\vec{E}_{\epsilon}|}{|\vec{E}_{\rm m}|}e^{i(\Delta\phi_{\rm m}+\phi_{\epsilon}+\pi/2)}, \label{eq.11}
\end{eqnarray}
where
\begin{align} 
f_1^{\rm I}(\theta)&=\frac{\cos\theta_{\epsilon}t_{1s}\cos\theta_{\rm m}r_{2s}+\sin\theta_{\epsilon}t_{1p}\sin\theta_{\rm m}r_{2p}}{\cos\theta_{\rm m}t_{1s}\cos\theta_{\rm m}r_{2s}+\sin\theta_{\rm m}t_{1p}\sin\theta_{\rm m}r_{2p}}\label{eq12}\\
f_2^{\rm I}(\theta)&=\frac{\cos\theta_{\epsilon}r_{1s}\cos\theta_{\rm m}t_{2s}+\sin\theta_{\epsilon}r_{1p}\sin\theta_{\rm m}t_{2p}}{\cos\theta_{\rm m}{r}_{1s}\cos\theta_{\rm m}t_{2s}+\sin\theta_{\rm m}{r_{1p}}\sin\theta_{\rm m}t_{2p}}\label{eq13}
\end{align}
are the coupling coefficients of the parasitic beam to each channel, which define how much the parasitic beam degrades the phase measurement. The superscripts $\mathrm{I}$ and, later on, $\mathrm{II}$ denote Type-I and Type-II
parasitic interference, respectively, as discussed in
Secs.~\ref{2.1} and~\ref{2.2}.

\begin{figure}[ht] 
\centering\includegraphics[width = .9\linewidth]{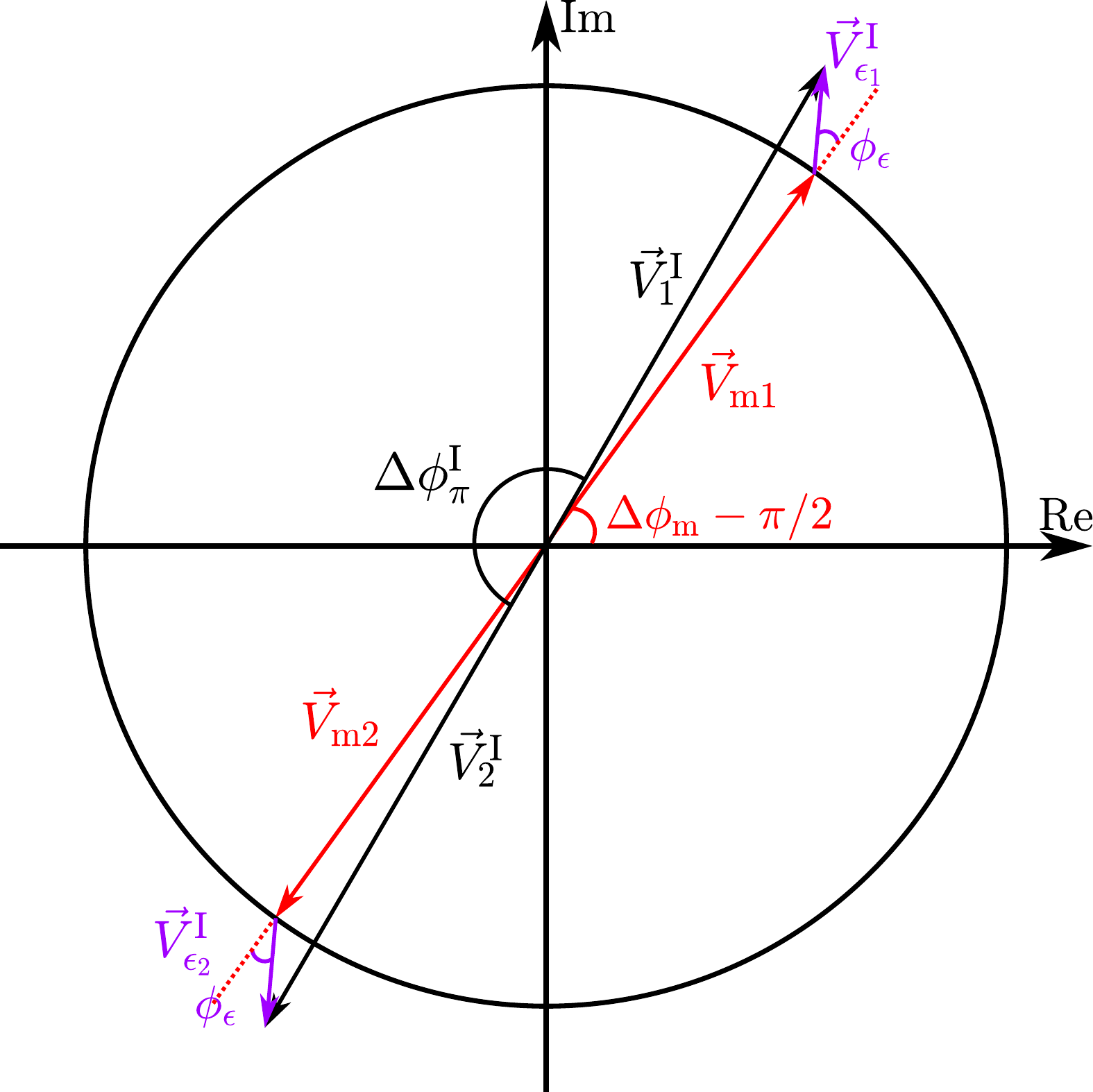}
\caption{Phase vectors in the presence of a type-I parasitic beam as measured at PD$_{1}$ and PD$_{2}$. $ \vec{V}_{\rm m_1}$ and $\vec{V}_{\rm m_2}$ are the two main phase vectors, which are antiparallel. $\vec{V}^{\rm I}_{\rm \epsilon_1}$ and $\vec{V}^{\rm I}_{\rm \epsilon_2}$ are the parasitic phase vectors, which are also antiparallel. Since the relative field amplitudes of the parasitic beam differ at each PD ($|\vec{V}^{\rm I}_{\rm \epsilon_1}|\ne|\vec{V}^{\rm I}_{\rm \epsilon_2}|$) due to the beamsplitter splitting ratios, the overall phase vectors ($\vec{V}^{\rm I}_{\rm 1},\vec{V}^{\rm I}_{\rm 2}$) measured at each channel are no longer antiparallel, and thus $\Delta\phi^{\rm I}_\pi=\angle \vec{V}^{\rm I}_{2}-\angle \vec{V}^{\rm I}_{1}\neq\pi$ unlike the ideal case where no parasitic beam is present.} \label{fig: V1}
\end{figure}

Figure~\ref{fig: V1} represents the phase vectors given by Eqs.~(\ref{eq.8}) and (\ref{eq.9}). Note that the parasitic phase vectors are antiparallel to each other. A fixed angle $\phi_\epsilon$ between the parasitic phase vectors and their corresponding main phase vectors is obtained, which we define as type-I parasitic interference and is indicated as superscript I in all equations. Assuming the parasitic beam is much weaker than the main beams, the phase difference between the phases measured at PD$_{1}$ and PD$_{2}$ is
\begin{equation}\label{eq.14}
 \Delta\phi^{\rm I}_{\pi} = \angle \vec{V}^{\rm I}_{2}-\angle \vec{V}^{\rm I}_{1}= \pi+[f_2^{\rm I}(\theta)-f_1^{\rm I}(\theta)]\frac{|\vec{E}_{\epsilon}|}{|\vec{E}_{\rm m}|}\sin(\phi_\epsilon),
\end{equation}
where the phase of the complex vector $\vec{V}_{i}$ is defined as
\begin{equation}\label{eq.add}
    \angle\vec{V}_{i}
    = \operatorname{atan2}
    \left(
    \operatorname{Im}(\vec{V}_{i}),
    \operatorname{Re}(\vec{V}_{i})
    \right).
\end{equation}
$\Delta\phi_{\pi}$ denotes the differential phase obtained from the $\pi$ test, which employs a cancellation principle analogous to that of the differential interferometry scheme introduced later. The second term in Eq.~(\ref{eq.14}) represents the parasitic contribution to the phase measurement. This contribution consists of three factors. The first is the coupling coefficient $f_2^{\rm I}(\theta)-f_1^{\rm I}(\theta)$, which describes the system's susceptibility to parasitic interference and depends on the polarization states of the input beams ($\theta_{\rm m}$, $\theta_{\rm \epsilon}$) and polarization-dependent system parameters such as the BS reflectivity and transmissivity, surface coatings, and losses. The second factor is the relative field amplitude of the parasitic beam with respect to the main beam. The third factor is the phase of the parasitic beam, which is proportional to the optical path length traveled by the parasitic beam, and is often referred to as non-linear optical path length (OPL) noise.

To mitigate the effect of parasitic beams, all three factors can be addressed. For the last two factors, common strategies include reducing the parasitic beam amplitude $|\vec{E}_{\epsilon}|$ and stabilizing its OPL $\phi_{\epsilon}$. The former is achieved through, e.g., antireflective (AR) coatings and wedged surfaces~\cite{Liepmann:92}, while the latter requires minimizing environmental disturbances (temperature, vibrations, pressure, etc.) that cause optical path length fluctuations and drifts over time. However, these approaches are often insufficient, particularly in the low-frequency regime (tens of millihertz and below), where controlling environmental conditions becomes especially challenging. In this paper, we address the first factor, the coupling coefficients [$f_{i}(\theta)$], through optimization of polarization direction, common-mode noise rejection configurations, and balanced detection. 

In the following sections, we extend the model of parasitic interference and define type-I and type-II parasitic interference. The former consists of anti-parallel parasitic vectors at PD$_{1}$ and PD$_{2}$ (see small purple vectors $\vec{V}^{\rm I}_{\epsilon_1}$ and $\vec{V}^{\rm I}_{\epsilon_2}$ in Fig.~\ref{fig: V1}), while in the latter the parasitic vectors are parallel. We model the effects of parasitic interference on $\pi$
tests and balanced detection. The $\pi$
test is analogous to a two-interferometer common-mode noise rejection scheme described in Sec.~\ref{S3}, making it a useful diagnostic for characterizing parasitic interference in such configurations.

\subsection{Type-I parasitic interference} \label{2.1}
\begin{figure}[ht] 
\centering\includegraphics[width = .9\linewidth]{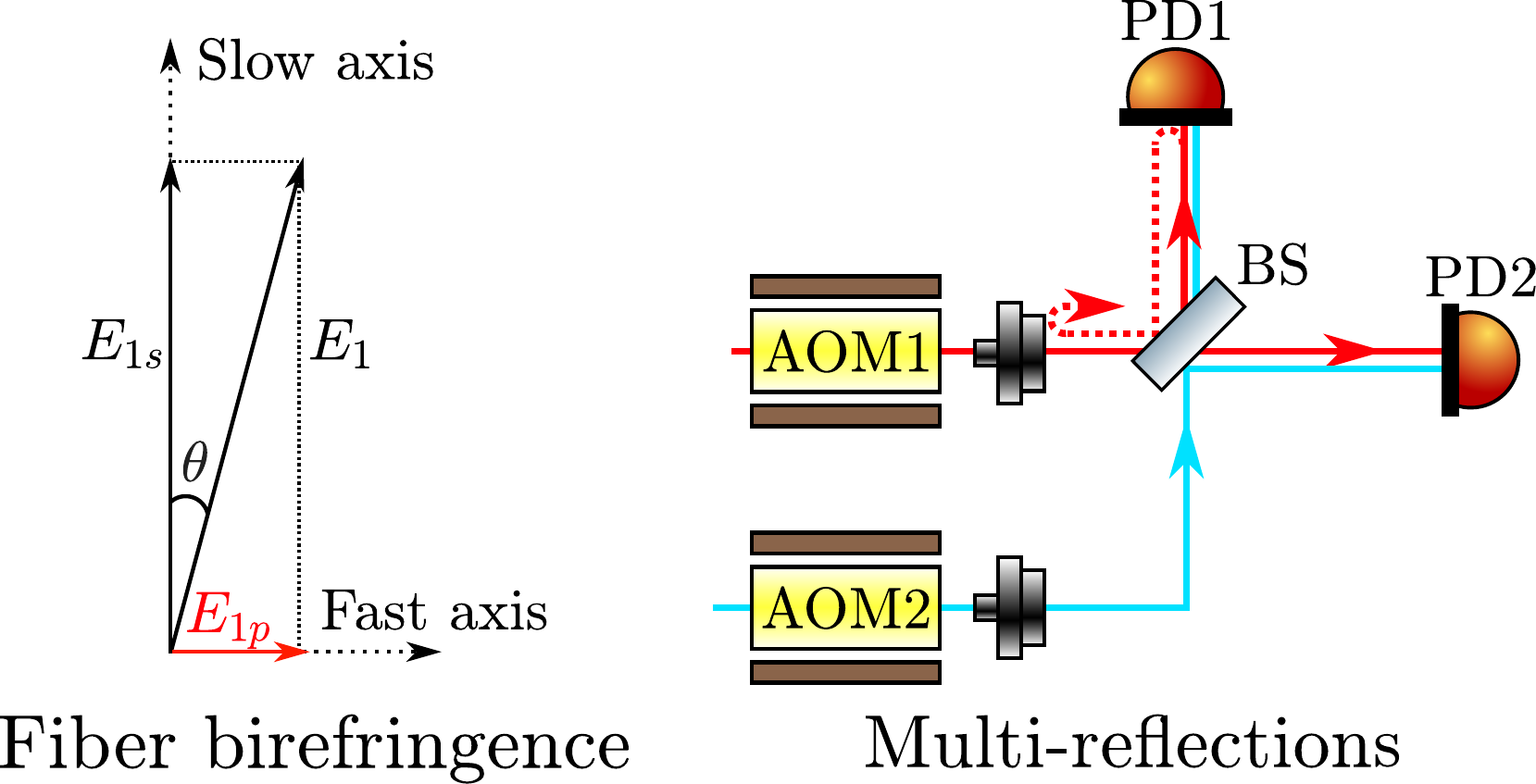}
\caption{Examples of type-I parasitic beams. Left: misalignment between the polarization direction of the input beam and the slow axis of the fiber, combined with the fiber birefringence effect, results in two orthogonally polarized beams with different phases. One is treated as the main beam, while the other is treated as a parasitic beam. Right: solid lines show the optical path of the main beams. The dashed line shows an example of a parasitic beam caused by multiple reflections at optical surfaces (the photodiode and the fiber end facet), acquiring additional phase along its path.}\label{fig: G1exp}
\end{figure}

Type-I parasitic interference originates from a parasitic beam entering the recombination BS from the same input and at the same frequency as the main beam at that input (i.e., input 1 at $\Omega_1$). In all phase channels (here in both PD1 and PD2), the angle between type-I parasitic phase vectors and their corresponding main phase vectors is the same ($\angle(\vec{V}^{\rm I}_{\epsilon_1},\vec{V}_{\rm{m}_1})=\angle(\vec{V}^{\rm I}_{\epsilon_2},\vec{V}_{\rm{m}_2})=\phi_\epsilon$). Type-I parasitic beams can arise from, e.g., multiple reflections from optical surfaces, optical birefringence (especially in optical fibers), and polarization drift. Figure~\ref{fig: G1exp} shows two examples of type-I parasitic beam generation: birefringence (left) and multiple back-reflections (right). For any polarization misalignment of the input beam, the OPL experienced by the two polarization states differs due to the birefringence, resulting in a phase-shifted parasitic beam. Ideally, the two polarization states are orthogonal, and no interference occurs. However, further polarization misalignment of polarization-dependent optics (e.g., polarizer, BS, and wave plate) can enable interference between them, contributing to the measured phase noise. This effect is greatly amplified in fiber-coupled systems due to the long fiber length and the refractive index difference between the fast and slow axes. Another possible source of parasitic beams is multiple reflections at optical surfaces (Fig.~\ref{fig: G1exp}, right). For instance, for AR-coated surfaces with reflectance of 0.5\%, two reflections give rise to a parasitic beam with a relative intensity of about $10^{-5}$ with respect to the main beam. 

Let us consider the setup shown in Fig.~\ref{IFO MZ} and carry out a $\pi$ test. The phase noise term in Eq.~(\ref{eq.14}) can be rewritten as 
\begin{align}
 \delta\phi^{\rm I}_{\pi} &=\Delta\phi^{\rm I}_{\pi}-\pi= [f_2^{\rm I}(\theta)-f_1^{\rm I}(\theta)]\frac{|\vec{E}_{\epsilon}|}{|\vec{E}_{\rm m}|}\sin\phi_{\epsilon}
 \nonumber\\
 &= f_{\Delta\phi }^{\rm I}(\theta)\frac{|\vec{E}_{\epsilon}|}{|\vec{E}_{\rm m}|}\sin\phi_{\epsilon},
\end{align}
where $f_{\Delta\phi }^{\rm I}$ denotes the coupling coefficient of type-I parasitic interference to the differential phase. A parasitic beam can be decomposed into two polarization components for ease of analysis. The perpendicular component can be filtered out by inserting a polarizer before the recombination BS, whereas the parallel component cannot be filtered out since it shares the polarization of the main beam. The coupling coefficients of the two components are
\begin{eqnarray}
 f_{\Delta\phi||}^{\rm I}&=&f_{\Delta\phi}^{\rm I}(\theta=\theta_{\rm m}) \label{eq.16} \\
 f_{\Delta\phi\perp}^{\rm I}&=&f_{\Delta\phi}^{\rm I}(\theta=\theta_{\rm m}+\pi/2) \label{eq.17},
\end{eqnarray}
where $||$ and $\perp$ indicate the parallel and perpendicular polarization components of the parasitic beam, respectively. The parallel component, Eq.~(\ref{eq.16}), is exactly zero, indicating that the interferometer is insensitive to the parallel component of a type-I parasitic beam. This occurs because the parasitic beam shares the same polarization as the main beam, so the amplitude ratio between them after the BS remains the same at both outputs. In the phase vectors picture, the two parasitic phase vectors have the same amplitude and angle with respect to the main phase vector, such that their contributions to the differential phase fully cancel out. 

\begin{figure}[ht] 
\centering\includegraphics[width = \linewidth]{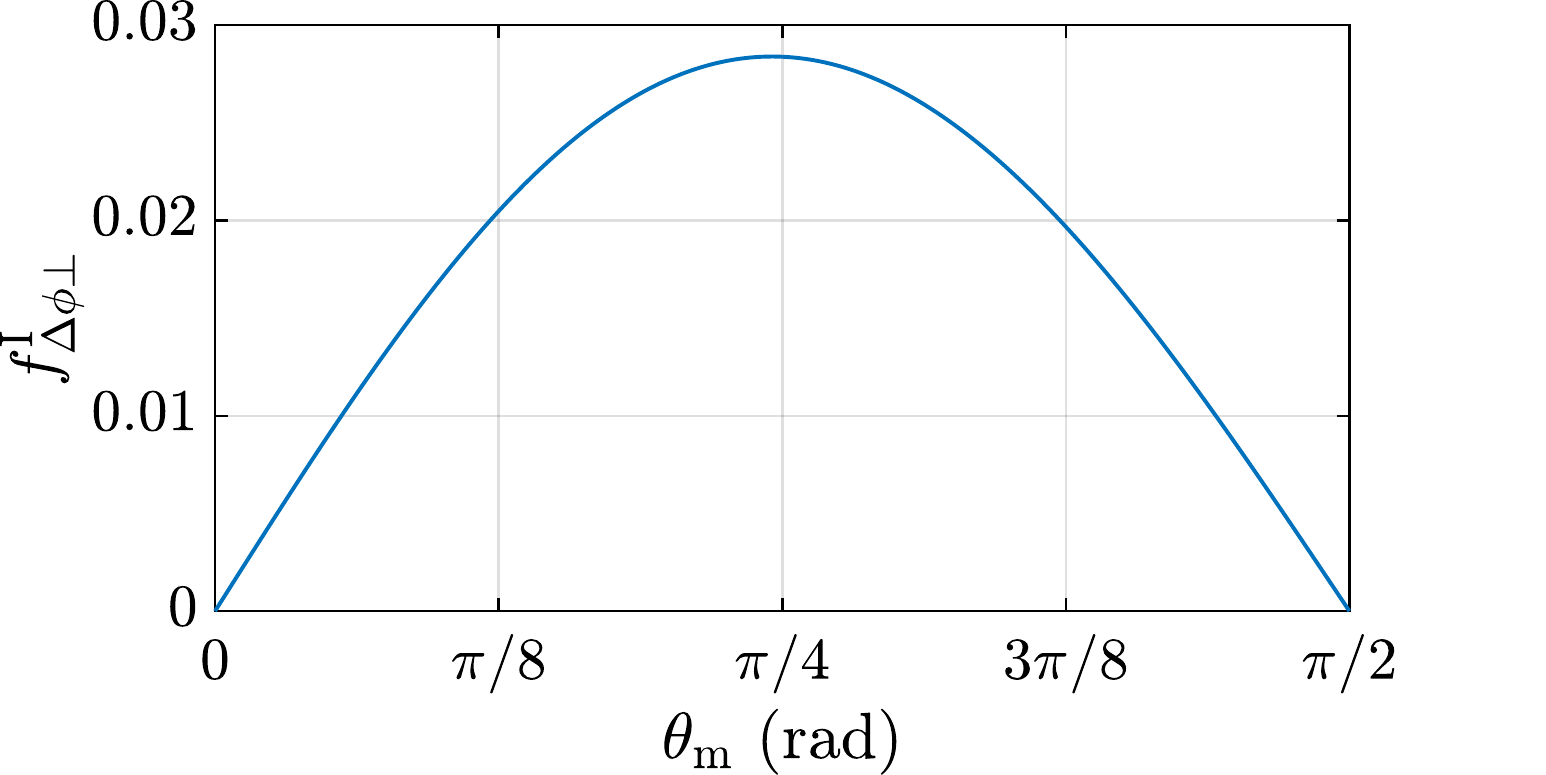}
\caption{Perpendicular component of type-I parasitic interference coupling coefficient ($f_{\Delta\phi\perp}^{\rm I}$) versus input polarization.} \label{fig: f1dpperp}
\end{figure}

The perpendicular component of the parasitic beam does not always cancel out, as shown in Fig.~\ref{fig: f1dpperp}, where Eq.~(\ref{eq.17}) is plotted in absolute values for different input polarization directions of the main beam $\theta_{\rm m}$. All coupling coefficients in this manuscript are plotted in absolute values unless stated otherwise, where the absolute value symbol is omitted. Intuitively, one might expect the perpendicular component to have no effect since it cannot interfere with the main beam. However, due to the imperfect splitting ratio of the recombination BS, the polarization of the output beam is rotated by a small angle with respect to the input polarization, enabling interference and the corresponding phase noise. For $\theta_{\rm m}=0$ or $\pi/2$, the beam polarizations are aligned with either axis of the BS, so no polarization rotation occurs, the parasitic beam does not interfere with the main beam, and no phase noise is present. For the interferometer operating with polarization oriented at about $\pi/4$, polarization rotation from the BS is maximized, yielding a maximum coupling coefficient of 0.028 for the BS parameters given at the beginning of this section. 

In summary, for type-I parasitic interference in the interferometer $\pi$ test, the parallel component does not contribute to phase noise, whereas the perpendicular component does, depending on the polarization orientation. The latter can be effectively mitigated by inserting linear polarizers before the recombination BS or by aligning the beam polarization with the BS axes.

\subsection{Type-II parasitic interference} \label{2.2}
\begin{figure}[b] 
\centering\includegraphics[width = .9\linewidth]{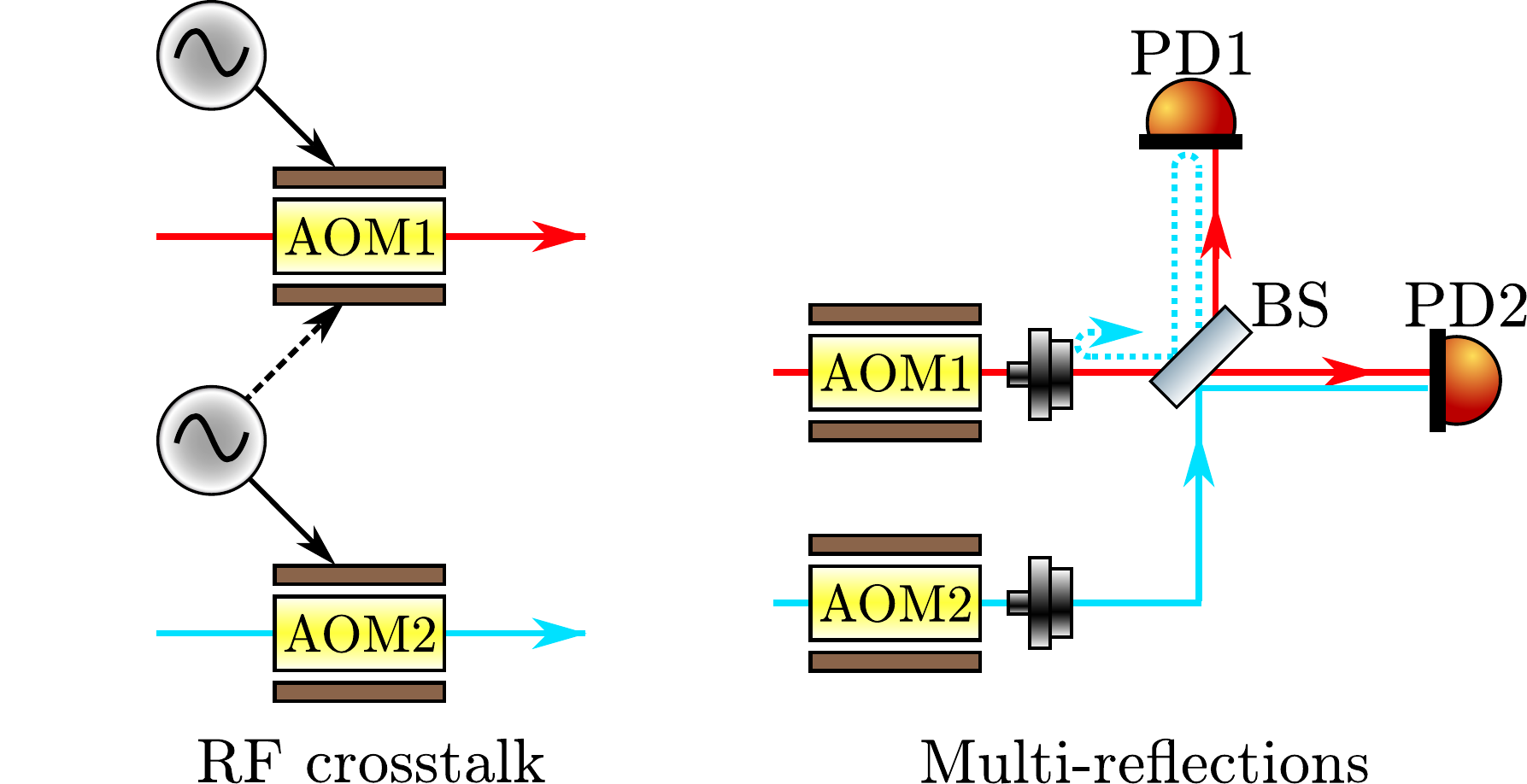}
\caption{Examples of type-II parasitic beams. Solid red and blue lines represent the two main beams. Dashed lines represent the RF and optical leakage causing the type-II parasitic beams, which are entering the BS from input 1 but with frequency $\Omega_2$.}\label{fig: G2exp}
\end{figure}

Type-II parasitic interference originates from a parasitic beam entering the recombination BS from one input while carrying the frequency of the main beam at the other input (i.e., entering from input 1 at frequency $\Omega_2$). In all phase channels, all type-II parasitic phase vectors are parallel to each other ($\vec{V}^{\rm II}_{\epsilon_i}~||~\vec{V}^{\rm II}_{\epsilon_j}$), as shown later in this section. Examples of type-II parasitic beam generation are RF crosstalk between
AOMs and multiple reflections from non-perfect optical components as shown in Fig.~\ref{fig: G2exp}. For RF crosstalk between AOMs, the parasitic beam intensity ratio depends on the RF isolation between the AOMs. A typical isolation measured in our setup is 50\,dB, which corresponds to an intensity ratio of $10^{-5}$ between the parasitic and the main beams. Similar to the type-I case, multiple reflections at AR-coated optical surfaces can produce a parasitic beam with a relative intensity of $10^{-5}$ with respect to the main beam.

A derivation analogous to that in the previous section yields the phase vectors and coupling coefficients for type-II parasitic interference. The resulting phase vectors are
\begin{align} 
\vec{V}^{\rm II}_1&=\vec{V}_{\rm m_1}+\vec{V}^{\rm II}_{\epsilon_1}=e^{i(\Delta\phi_{\rm m}-\pi/2)}+f_1^{\rm II}(\theta)\frac{|\vec{E}_{\epsilon}|}{|\vec{E}_{\rm m}|}e^{i\phi_{\epsilon}}\label{eq18}\\
\vec{V}^{\rm II}_2&=\vec{V}_{\rm m_2}+\vec{V}^{\rm II}_{\epsilon_2}=e^{i(\Delta\phi_{\rm m}+\pi/2)}+f_2^{\rm II}(\theta)\frac{|\vec{E}_{\epsilon}|}{|\vec{E}_{\rm m}|}e^{i\phi_{\epsilon}},\label{eq19}
\end{align}
where
\begin{align} 
f_1^{\rm II}(\theta)&=\frac{\cos\theta_{\epsilon1}t_{1s}\cos\theta_{\rm m}t_{1s}+\sin\theta_{\epsilon}t_{1p}\sin\theta_{\rm m}t_{1p}}{\cos\theta_{\rm m}t_{1s}\cos\theta_{\rm m}r_{2s}+\sin\theta_{\rm m}t_{1p}\sin\theta_{\rm m}r_{2p}}\label{eq20}\\
f_2^{\rm II}(\theta)&=\frac{\cos\theta_{\epsilon1}r_{1s}\cos\theta_{\rm m}r_{1s}+\sin\theta_{\epsilon}r_{1p}\sin\theta_{\rm m}r_{1p}}{\cos\theta_{\rm m}r_{1s}\cos\theta_{\rm m}t_{2s}+\sin\theta_{\rm m}r_{1p}\sin\theta_{\rm m}t_{2p}}.\label{eq21}
\end{align}
Note that compared to Eqs.~(\ref{eq.10}) and (\ref{eq.11}), the $\pi/2$ terms in the parasitic phase in Eqs.~(\ref{eq18}) and (\ref{eq19}) vanish. This is because the interference at the heterodyne frequency occurs between the parasitic beam and the main beam~1, which share a common phase shift along the optical path.

\begin{figure}[!htbp] 
\centering\includegraphics[width=.9\linewidth]{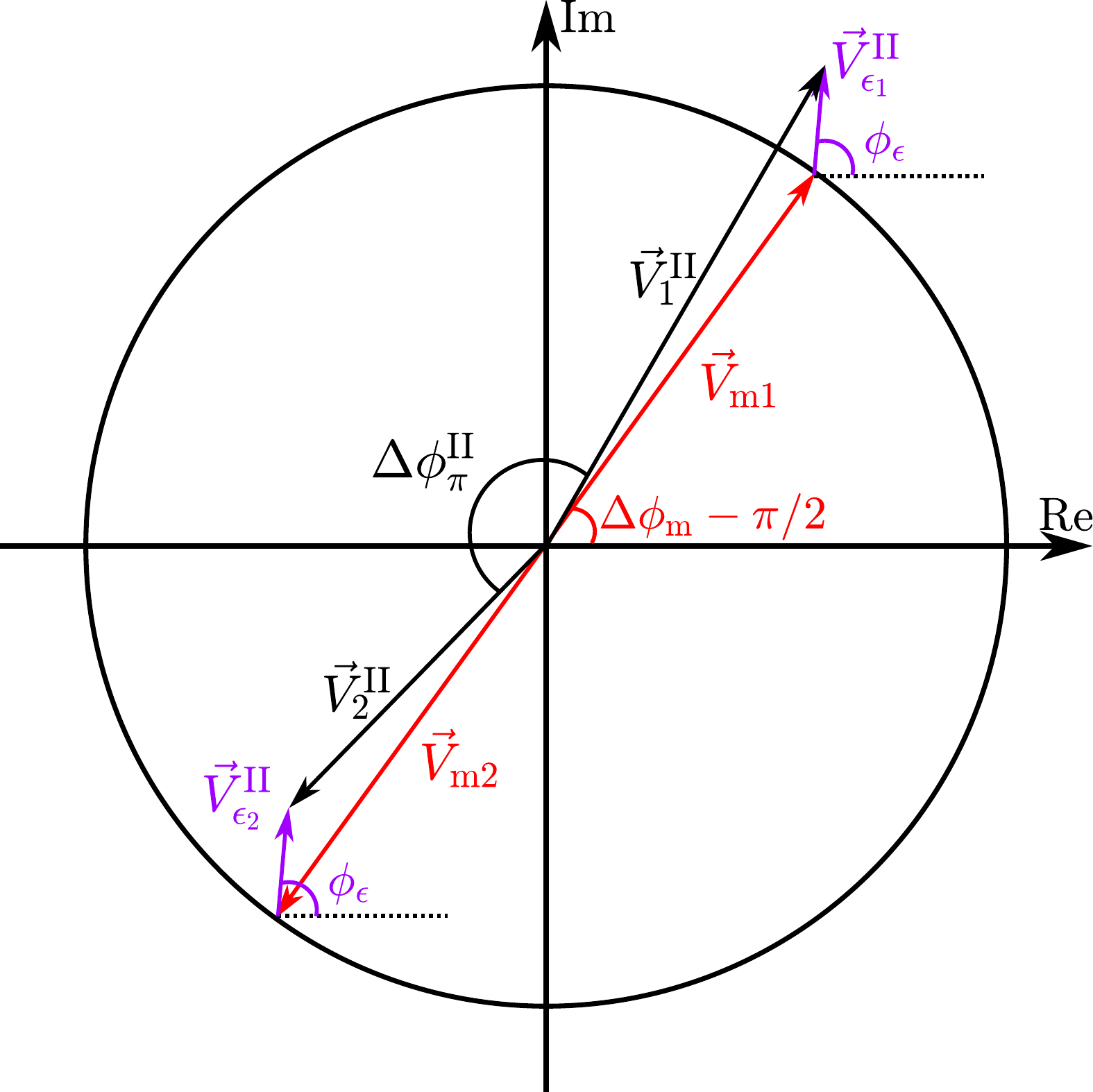}
\caption{Phase vectors in the presence of a type-II parasitic beam. $\vec{V}_{\rm m_1}$ and $\vec{V}_{\rm m_2}$ are the two main phase vectors, which are antiparallel. $\vec{V}^{\rm II}_{\rm \epsilon_1}$ and $\vec{V}^{\rm II}_{\rm \epsilon_2}$ are the phase vectors caused by type-II parasitic interference, which are parallel. The overall phase vectors ($\vec{V}^{\rm II}_{\rm 1},\vec{V}^{\rm II}_{\rm 2}$) measured at PD$_{1}$ and PD$_{2}$ are clearly not antiparallel, preventing $\Delta\phi^{\rm II}_\pi=\angle \vec{V}^{\rm II}_{1}-\angle \vec{V}^{\rm II}_{2}$ from equaling $\pi$.} \label{fig: V2}
\end{figure}

Figure~\ref{fig: V2} shows the phase vectors acquired by the two PDs. Here, both parasitic interference phase vectors (purple small vectors, $\vec{V}_{\epsilon_1}^{\rm II}$ and $\vec{V}_{\epsilon_2}^{\rm II}$) are parallel. The phase difference is
\begin{align}
 \Delta\phi^{\rm II}_{\pi} &= \angle \vec{V}^{\rm II}_{2} - \angle \vec{V}^{\rm II}_{1} \nonumber\\
 &= \pi -[f_2^{\rm II}(\theta)+f_1^{\rm II}(\theta)]\frac{|\vec{E}_{\epsilon}|}{|\vec{E}_{\rm m}|}\sin\left[\phi_{\epsilon}-(\Delta\phi_{\rm m}-\pi/2)\right]
 \nonumber\\
 &=\pi- f_{\Delta\phi}^{\rm II}(\theta)\frac{|\vec{E}_{\epsilon}|}{|\vec{E}_{\rm m}|}\sin\left[\phi_{\epsilon}-(\Delta\phi_{\rm m}-\pi/2)\right]
 ,\label{eq.22}
\end{align}
where it is readily seen that the parasitic phase noise now depends not only on $\phi_{\epsilon}$ but also on $\Delta\phi$, meaning that the type-II phase noise contribution varies with the measurement target too. In addition, the two parasitic phase terms add constructively, such that type-II parasitic interference couples more strongly than type-I for the same parasitic beam amplitude. 

\begin{figure}[htbp] 
\centering\includegraphics[width = \linewidth]{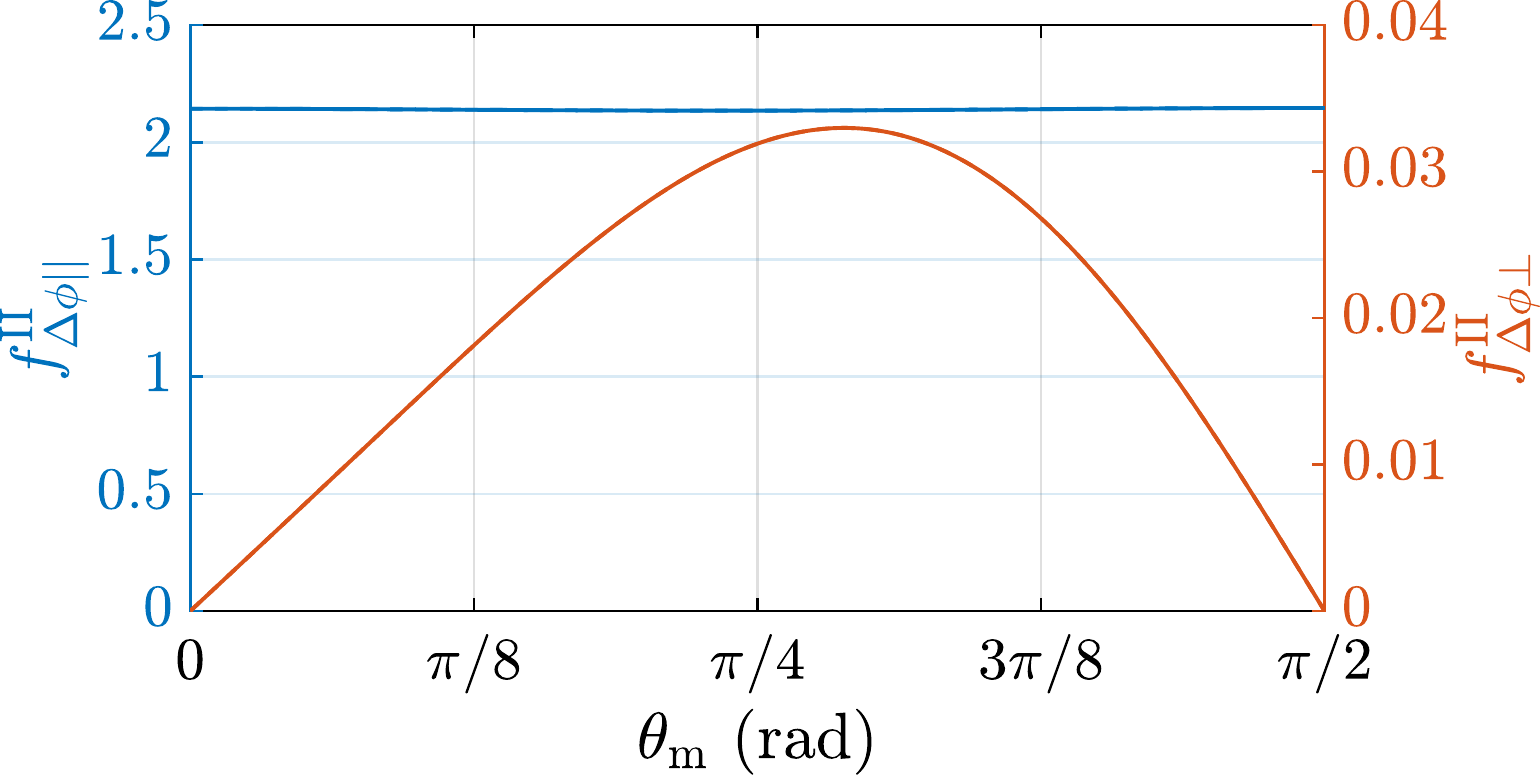}
\caption{Type-II parasitic interference coupling coefficient versus input polarization. Left y-axis: parallel component. Right y-axis: perpendicular component. Note that for an ideal recombination BS, the left plot remains unchanged, while the right panel is identically zero for all $\theta_{\rm m}$.} \label{fig: f2dp}
\end{figure}

As in Sec.~\ref{2.1}, we decompose the coupling coefficient into parallel and perpendicular components. Figure~\ref{fig: f2dp} shows the coefficients of type-II parasitic interference as a function of the main beam polarization orientation $\theta_{\rm m}$ and for the BS presented at the beginning of the section. The parallel component has a coupling coefficient of approximately 2.14 regardless of $\theta_{\rm m}$, indicating that a strong noise contribution is expected. For the perpendicular component, minimum noise coupling occurs when the beam polarization is aligned with either axis of the BS, as in the type-I case. The maximum coupling coefficient is 0.033, which is of the same order as the type-I case. 

In summary, for type-II parasitic interference in the $\pi$
test, the parallel component contributes strongly to phase noise regardless of the polarization orientation, whereas the perpendicular component vanishes for $\theta_{\rm m}=0$ and $\pi/2$, and is maximized near $\theta_{\rm m}=1~\rm rad$. The latter, while small, can be further reduced by inserting linear polarizers before the recombination BS. 

\subsection{Balanced detection} \label{S2.3}
In Secs.~\ref{2.1} and~\ref{2.2}, the coupling coefficients of type-I and type-II parasitic interference have been investigated for the $\pi$ test. The perpendicular components have small coupling coefficients, which can be further mitigated by inserting polarizers before the BS. However, the parallel components share the polarization direction of the main beams, and the coupling from type-II parasitic interference is particularly strong (cf. Fig.~\ref{fig: f2dp}). As shown in Fig.~\ref{fig: V2}, the parasitic phase vectors at the two channels are parallel. Thus, a balanced detection scheme in which the phase vectors are subtracted can reduce the coupling of type-II parasitic interference, while maintaining the phase of interest.

We now consider the setup shown in Fig.~\ref{IFO MZ} and apply balanced
detection by subtracting the phase vectors obtained at the two output ports
of the BS. Here, the goal is to measure $\Delta\phi_{\rm m}$ ($=\phi_{\rm m_1}-\phi_{\rm m_2}$) and characterize the phase noise contribution from both types of parasitic interference. Consider the parallel component (the perpendicular component is assumed negligible if linear polarizations are inserted) of the parasitic beam and perform vector subtraction in the complex plane. The resulting phases with the presence of either type of parasitic beam are 
\begin{eqnarray}
 \phi_{\rm BD}^{\rm I}&=&\angle(\vec{V}^{\rm I}_1-\vec{V}^{\rm I}_2) \nonumber\\
 &\simeq& \Delta\phi_{\rm m}-\frac{\pi}{2}+f_{\rm BD||}^{\rm I}\frac{|\vec{E}_{\epsilon}|}{|\vec{E}_{\rm m}|}\sin(\phi_{\epsilon}) \label{eq.23} \\
 \phi_{\rm BD}^{\rm II}&=&\angle(\vec{V}^{\rm II}_1-\vec{V}^{\rm II}_2) \nonumber\\
 &\simeq& \Delta\phi_{\rm m}-\frac{\pi}{2}+f_{\rm BD||}^{\rm II}\frac{|\vec{E}_{\epsilon}|}{|\vec{E}_{\rm m}|}\sin\left [\phi_{\epsilon}-\left(\Delta\phi_{\rm m}-\frac{\pi}{2}\right)\right ] , \nonumber \\
 \label{eq.24}
\end{eqnarray}
where
\begin{align}
 f_{\rm BD||}^{\rm I} &= \frac{f_1^{\rm I}(\theta_{\rm m})+f_2^{\rm I}(\theta_{\rm m})}{2} \label{eq.25} \\
 f_{\rm BD||}^{\rm II} &= \frac{f_1^{\rm II}(\theta_{\rm m})-f_2^{\rm II}(\theta_{\rm m})}{2}.\label{eq.26}
\end{align}
are the corresponding coupling coefficients for either type of parasitic interference. In Fig.~\ref{fig: V2}, in the presence of type-II parasitic interference, the amplitudes of the measured phase vectors ($\vec{V}^{\rm II}_1,~\vec{V}^{\rm II}_2$) differ slightly because the parasitic vectors are parallel while the main phase vectors are antiparallel. Consequently, the two parasitic phase vectors cannot fully cancel in balanced detection. For type-I, balanced detection offers no benefit since the parasitic vectors are anti-parallel (cf. Fig.~\ref{fig: V1}). Nevertheless, Eqs.~(\ref{eq.23}) and (\ref{eq.24}) are valid when the parasitic beam amplitudes are much smaller than the main beam amplitude, as is typically the case. 

\begin{figure}[ht]
\centering
\includegraphics[width=.9\linewidth]{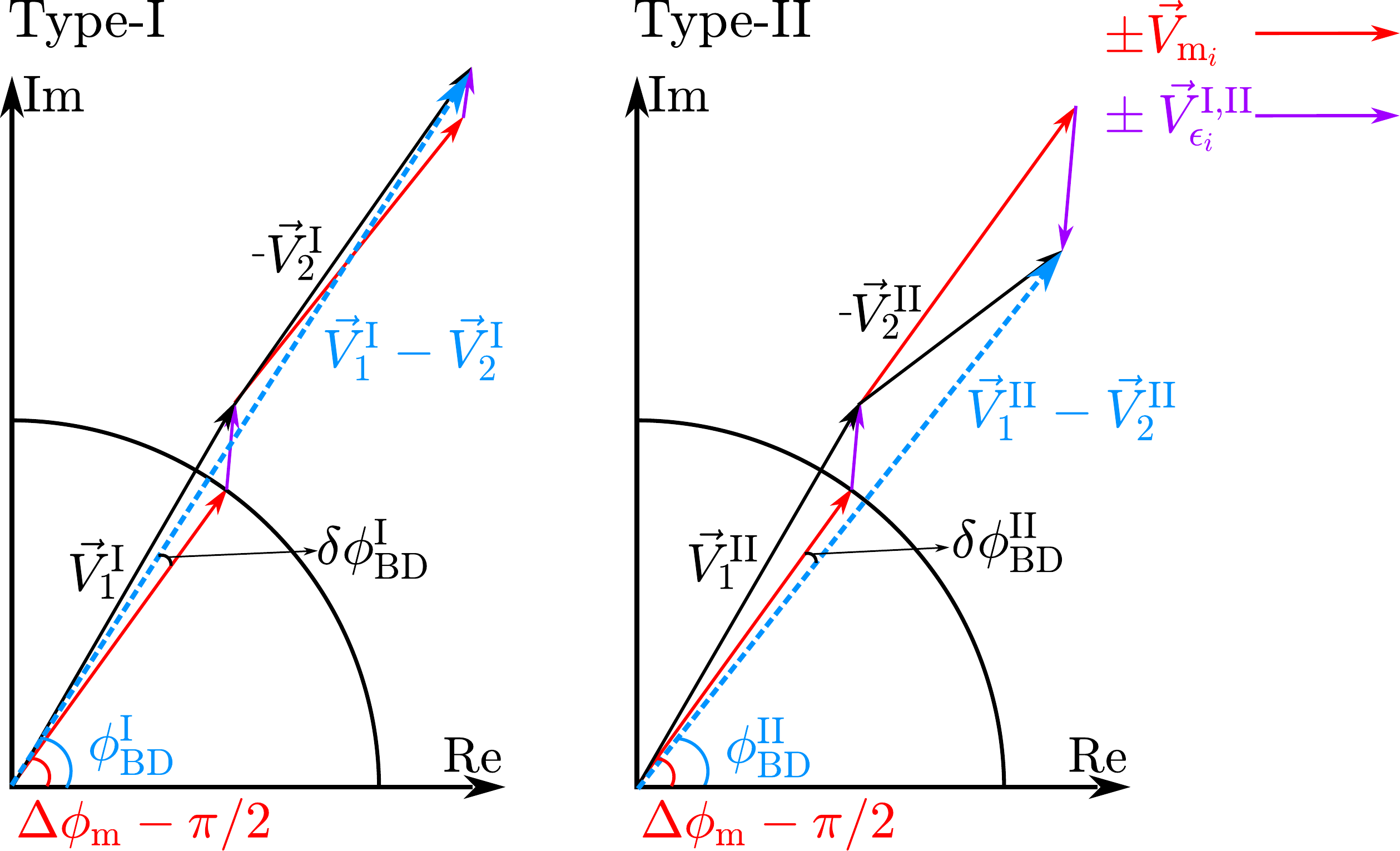}
\caption{Phase-vector representation of balanced detection in the presence
of type-I (left) and type-II (right) parasitic interference. The resulting phases
$\Delta\phi^{\rm I,II}_{\rm BD}$ are obtained by subtracting the phase
vectors measured by the two channels.}
\label{fig: Vbd}
\end{figure}

Figure~\ref{fig: Vbd} shows the phase vector representation of balanced
detection in the presence of either type of parasitic interference. The
resulting residual phase errors, $\delta\phi^{\rm I,II}_{\rm BD}$,
correspond to the last terms in Eqs.~(\ref{eq.23}) and~(\ref{eq.24}).

\begin{figure}[ht] 
\centering
\includegraphics[width = \linewidth]{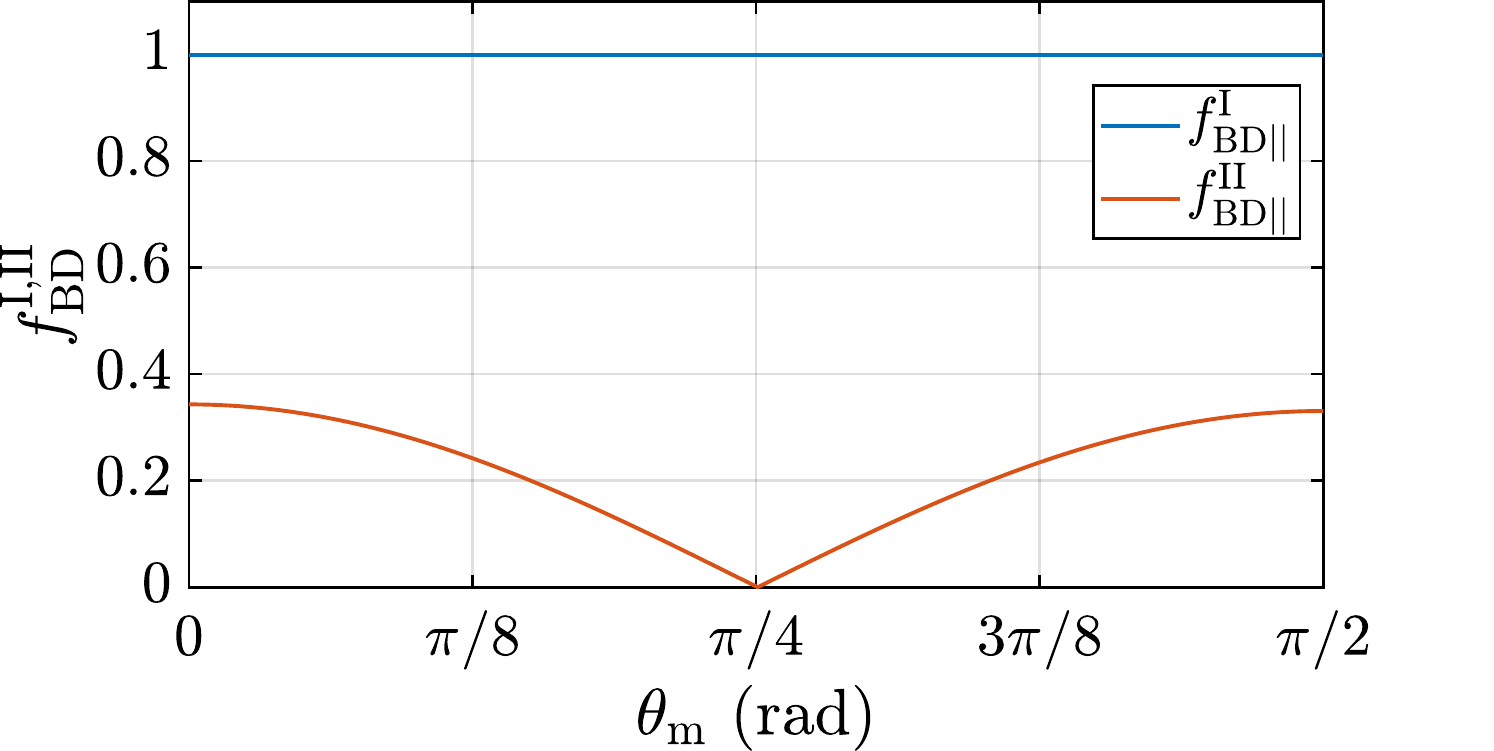}
\caption{Parasitic interference coupling coefficients of the parallel components versus input polarization direction in the balanced detection scheme.} \label{fig: fBDpi}
\end{figure}

Figure~\ref{fig: fBDpi} shows the parallel component coupling coefficients of parasitic interference versus input polarization for the balanced detection scheme. For type-I parasitic interference, the coupling coefficient is always unity regardless of polarization direction, confirming that balanced detection does not mitigate it. The maximum coupling coefficient of type-II parasitic interference is 0.33 at $\theta_{\rm m}=0\rm~{and}~\pi/2$, which is a factor of six lower than in the $\pi$ test. Note that for an input polarization of $\pi$/4, the transmittance and reflectance of the BS are equal, leading to equal parasitic beam amplitudes at the two PDs, resulting in further cancellation. 

Table~\ref{tab:symbols} summarizes the main symbols used in the
parasitic-interference model.

\begin{table}[h]
\caption{Summary of the main symbols used in the parasitic-interference model.}
\label{tab:symbols}
\begin{ruledtabular}
\begin{tabular}{cl}
Symbol & Definition \\
\hline
$\vec{E}_{\rm m_\textit{i}}$        & Main optical field \\
$\vec{E}_{\epsilon}$     & Parasitic optical field \\
$\phi_{\rm m_\textit{i}}$           & Main-field phase \\
$\phi_{\epsilon}$        & Parasitic-field phase \\
$\vec{V}_{i}$            & Complex phase vector at detection channel $i$ \\
$f$                      & Parasitic interference coupling coefficient \\
$\Delta\phi_{\rm m}$     & Main interferometric phase \\
$\Delta\phi_{\pi}$       & Differential phase obtained from the $\pi$ test \\
$\phi_{\rm BD}$    & Phase obtained from vector subtraction in\\ &balanced detection\\
$\delta\phi$             & Phase error relative to the ideal case \\
Superscript $\mathrm{I}$ & Type-I parasitic interference \\
Superscript $\mathrm{II}$& Type-II parasitic interference \\
\end{tabular}
\end{ruledtabular}
\end{table}

\subsection{Experimental verification} \label{s2.4}
The coupling of parasitic interference to the phase measurement has been analyzed for the interferometer $\pi$ test and balanced detection in the previous sections. To perform experimental verification, a summing amplifier is added to the AOM~1 driver to introduce a controlled signal that generates an artificial parasitic beam with known and adjustable parameters (phase and amplitude). The RF driving signal of AOM~1 is
\begin{align}
 S_{\rm AOM_1} = S_{\rm m}\sin{(\Omega_1t+\phi_{\rm m_1})}+S_\epsilon\sin{(\Omega_\epsilon t+\phi_{\rm m_1}+\phi_\epsilon)},
\end{align}
where the first term corresponds to one of the main beams, and the second to the parasitic beam. The resulting electric field after AOM 1 is
\begin{align}
 \vec{E_1} = \vec{E}_{\rm m} e^{i(\Omega_1t+\phi_{\rm m_1})}+\vec{E}_{\epsilon}e^{i(\Omega_\epsilon t+\phi_{\rm m_1}+\phi_\epsilon)},
\end{align}
where both the amplitude and phase of the artificial parasitic beam can be controlled via the RF signal. For type-I parasitic beams, $\Omega_\epsilon=\Omega_1$, whereas for type-II, $\Omega_\epsilon=\Omega_2$. In this section, only the parallel component of the parasitic beam is considered, which is achieved by placing polarizers after the AOMs. These polarizers also ensure high polarization extinction ratios (PER) for both main beams. Three factors that appear in the equations introduced in the previous sections are: the polarization direction ($\theta$), the relative parasitic beam amplitude ($|\vec{E}_{\epsilon}|/|\vec{E}_{\rm m}|$), and the phase of the parasitic beam ($\phi_\epsilon$). The two types of parasitic phase noise have the general form:
\begin{eqnarray}
 \delta\phi_i^{\rm I} &=& f^{\rm I}_i(\theta)\frac{|\vec{E}_{\epsilon}|}{|\vec{E}_{\rm m_2}|}\sin(\phi^{\rm I}_\epsilon) \label{eq.29} \\
 \delta\phi_i^{\rm II} &=& f^{\rm II}_i(\theta)\frac{|\vec{E}_{\epsilon}|}{|\vec{E}_{\rm m_1}|}\sin\left[\phi^{\rm II}_\epsilon-\left(\Delta\phi_{\rm m}-\frac{\pi}{2}\right)\right],\label{eq.30}
\end{eqnarray} 
where subscript $i$ indicates the type of measurement: differential measurement $\Delta \phi$, or balanced detection $\phi_{\rm BD}$.

\subsubsection{Parasitic phase $\phi_{\epsilon}$}
\begin{figure}[ht]
\centering\includegraphics[width = \linewidth]{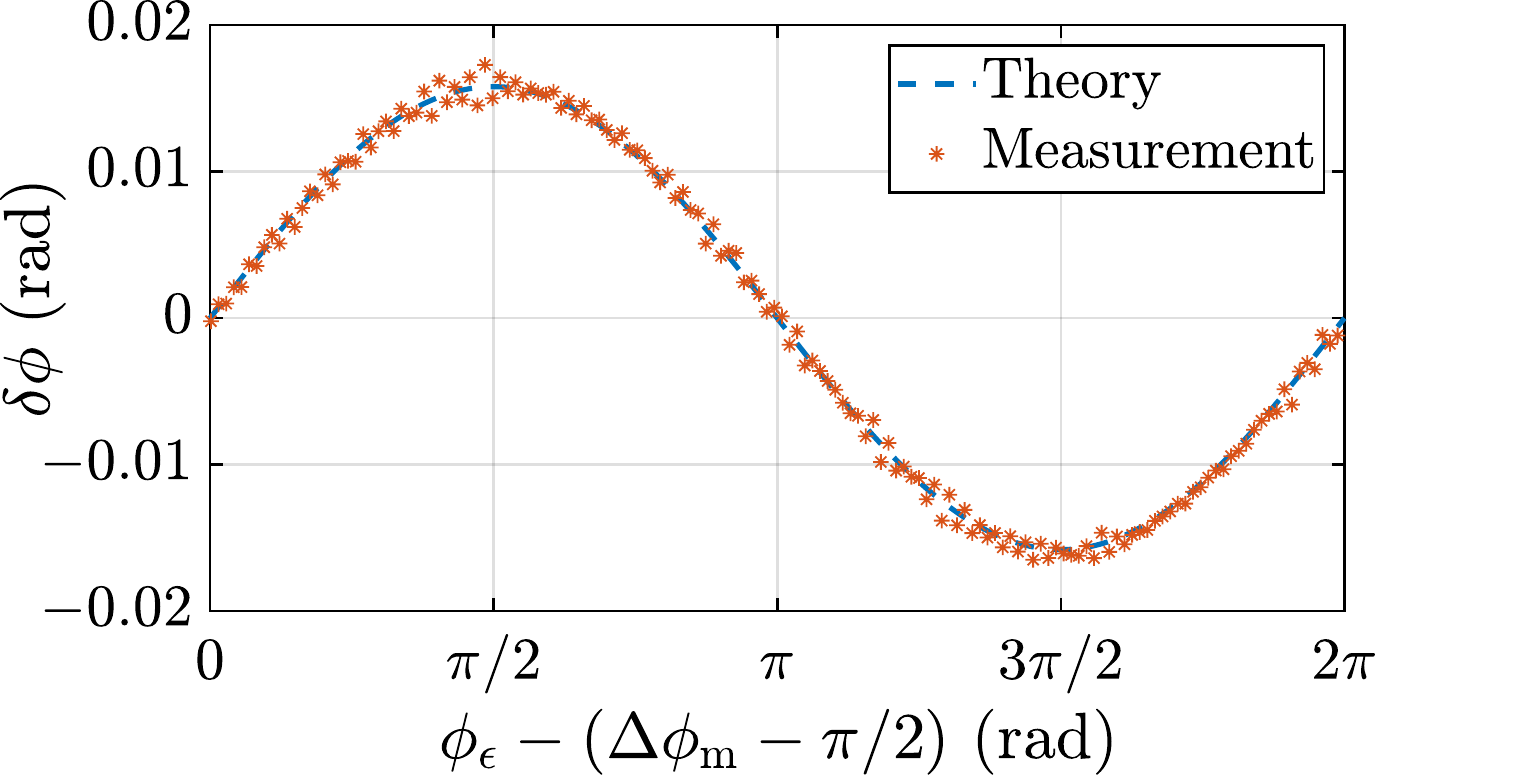}
\caption{Differential phase response ($\pi$ test) to the phase of a type-II parasitic beam ($\theta_\epsilon=\theta_{\rm m}\simeq0$, $|\vec{E}_{\epsilon}/\vec{E}_{\rm m}|^2=10^{-4}$).} \label{fig: pi_phi_vs_phie}
\end{figure}

First, we evaluate the term $\phi_\epsilon$ and its effect on the differential phase in a $\pi$ test. To this end, we inject a type-II parasitic beam with $\theta_{\rm m}\simeq0$ and $|\vec{E}_{\epsilon}/\vec{E}_{\rm m}|^2\simeq10^{-4}$, while sweeping $\phi_{\epsilon}$ over 2$\pi$. Simultaneously, we record the differential phase error ($\delta\phi = \Delta\phi_\pi-\pi$), which is plotted as a function of $\phi_{\epsilon}$ in Fig.~\ref{fig: pi_phi_vs_phie} together with the theoretical prediction from Eq.~(\ref{eq.22}). 
Good agreement between the model and measurements is observed. This result illustrates the nonlinear coupling between the parasitic phase and the main phase measurement. Since the coupling follows as a sinusoidal response, a slowly drifting parasitic phase upconverts low-frequency OPL fluctuations into phase noise at frequencies proportional to the rate of change of $\phi_{\epsilon}$. In particular, a constant phase drift at rate $\dot{\phi}_{\epsilon}$ produces noise concentrated at frequencies near $\dot{\phi}_{\epsilon}/2\pi$. For small fluctuations of $\phi_{\epsilon}$, the resulting differential phase noise depends on the mean value of $\phi_{\epsilon}$. 
 
\subsubsection{Polarization direction $\theta_{\rm m}$}
Next, we test the effect of beam polarization direction in both a $\pi$ test and a balanced detection configuration. Here we again set $|\vec{E}_{\epsilon}/\vec{E}_{\rm m}|^2\simeq10^{-4}$ and scan $\theta_{\rm m}$ from 0 to $2\pi$ in discrete steps of approximately 4$^{\rm o}$ or 8$^{\rm o}$ using a half-wave plate. At each polarization orientation, a phase modulation is applied to the parasitic beam. The modulation consists of a 10\,Hz sine wave with an amplitude of $\pi/6$\,rad around the main phase vector $\vec{V}_{1}$:
\begin{eqnarray}
 \phi^{\rm I}_{\epsilon}(t) &=&\frac{\pi}{6}\sin(2\pi 10t) \label{eq.31} \\ 
 \phi^{\rm II}_{\epsilon}(t) &=&\frac{\pi}{6}\sin(2\pi 10t)+\Delta\phi_{\rm m}-\pi/2. \label{eq.32}
\end{eqnarray}
In other words, the parasitic vector $\vec{V}^{\rm I,II}_{\epsilon_1}$ in Fig.~\ref{fig: V1} (and Fig.~\ref{fig: V2}) is initially parallel to the main phase vector $\vec{V}_{\rm m_1}$, and rotates around it with an amplitude of $\pi/6$\,rad. The parasitic vector $\vec{V}_{\epsilon_2}$ is parallel to $\vec{V}_{\rm m_2}$ for type-I and antiparallel for type-II. The particular amplitude is chosen to work in the semi-linear region of the sine function with a peak-to-peak value of unity. Plugging Eqs.~(\ref{eq.31}) and (\ref{eq.32}) into Eqs.~(\ref{eq.29}) and (\ref{eq.30}), the induced phase noise simplifies to the same form for both types of parasitic interference:
\begin{equation}
 \delta\phi_i^{\rm I, II} \simeq f^{\rm I, II}_i(\theta)\frac{|\vec{E}_{\epsilon}|}{|\vec{E}_{\rm m}|}\frac{1}{2}\sin(2\pi10t),
\end{equation}
which oscillates at 10\,Hz with a peak-to-peak phase noise amplitude ($\delta\phi^{\rm I, II}_{\rm pp}$) of
\begin{equation}
 \delta\phi_{\rm pp}^{\rm I, II} = f^{\rm I, II}_i(\theta)\frac{|\vec{E}_{\epsilon}|}{|\vec{E}_{\rm m}|}.
\end{equation}

\begin{figure}[b]
\centering\includegraphics[width = \linewidth]{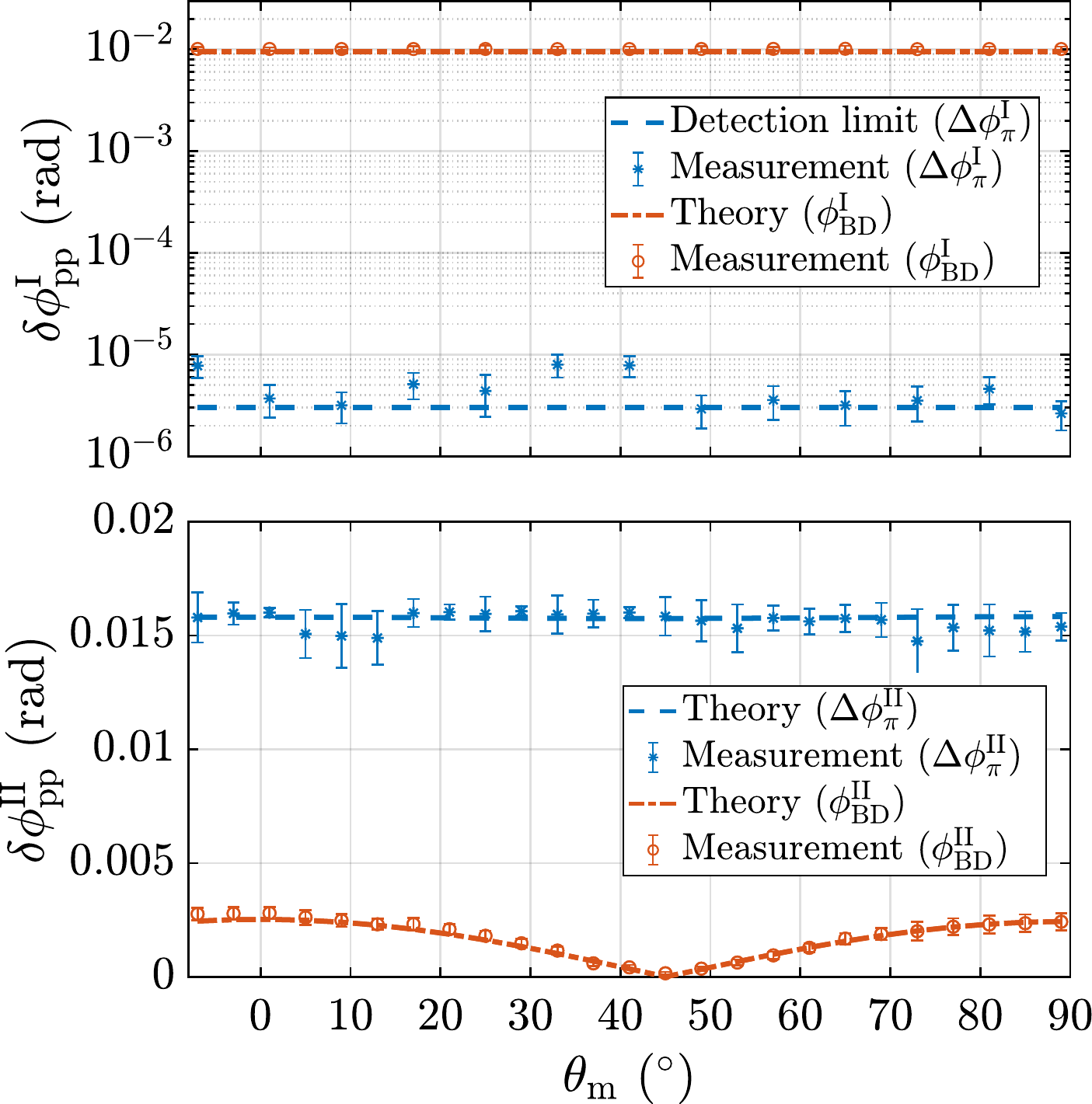}
\caption{Phase noise peak-to-peak amplitude versus input polarization direction in $\pi$ test ($\Delta\phi^{\rm I,II}_\pi$) and BD ($\phi^{\rm I,II}_{\rm BD}$) schemes. Top: type-I parasitic beam. Bottom: type-II parasitic beam. (Dashed lines: theoretical prediction. Markers: measured results with corresponding uncertainties.) \label{fig: pi_phi_vs_theta}}
\end{figure}

With the modulation applied, one-minute measurements are recorded for each configuration. The experimental peak-to-peak amplitude $\delta\phi^{\rm I, II}_{\rm pp}$ at 10\,Hz is shown in Fig.~\ref{fig: pi_phi_vs_theta} as a function of the input polarization direction $\theta_{\rm m}$ for both the $\pi$ test (blue traces) and the balanced detection configuration (red traces). For the $\pi$ test, type-I parasitic interference (top panel) does not contribute to the differential phase, and the results are effectively zero regardless of the input polarization direction, limited only by measurement uncertainty. However, type-II parasitic interference (bottom panel) exhibits a peak-to-peak phase amplitude of 16\,mrad. In balanced detection, type-I interference induces $\delta\phi^{\rm I}_{\rm pp}$ of 10\,mrad regardless of polarization direction, while type-II shows a more complex dependence: it reaches a maximum of 2.4\,mrad at 0$^ {\rm o}$ and 90$^{\rm o}$, and a null at 45$^{\rm o}$. Theoretical predictions are shown as dashed lines. Good agreement between the experimental results and the theoretical models is evident. 

\subsubsection{Parasitic beam amplitude $|\vec{E}_{\rm \epsilon}|$}
\begin{figure}[ht]
\centering\includegraphics[width = \linewidth]{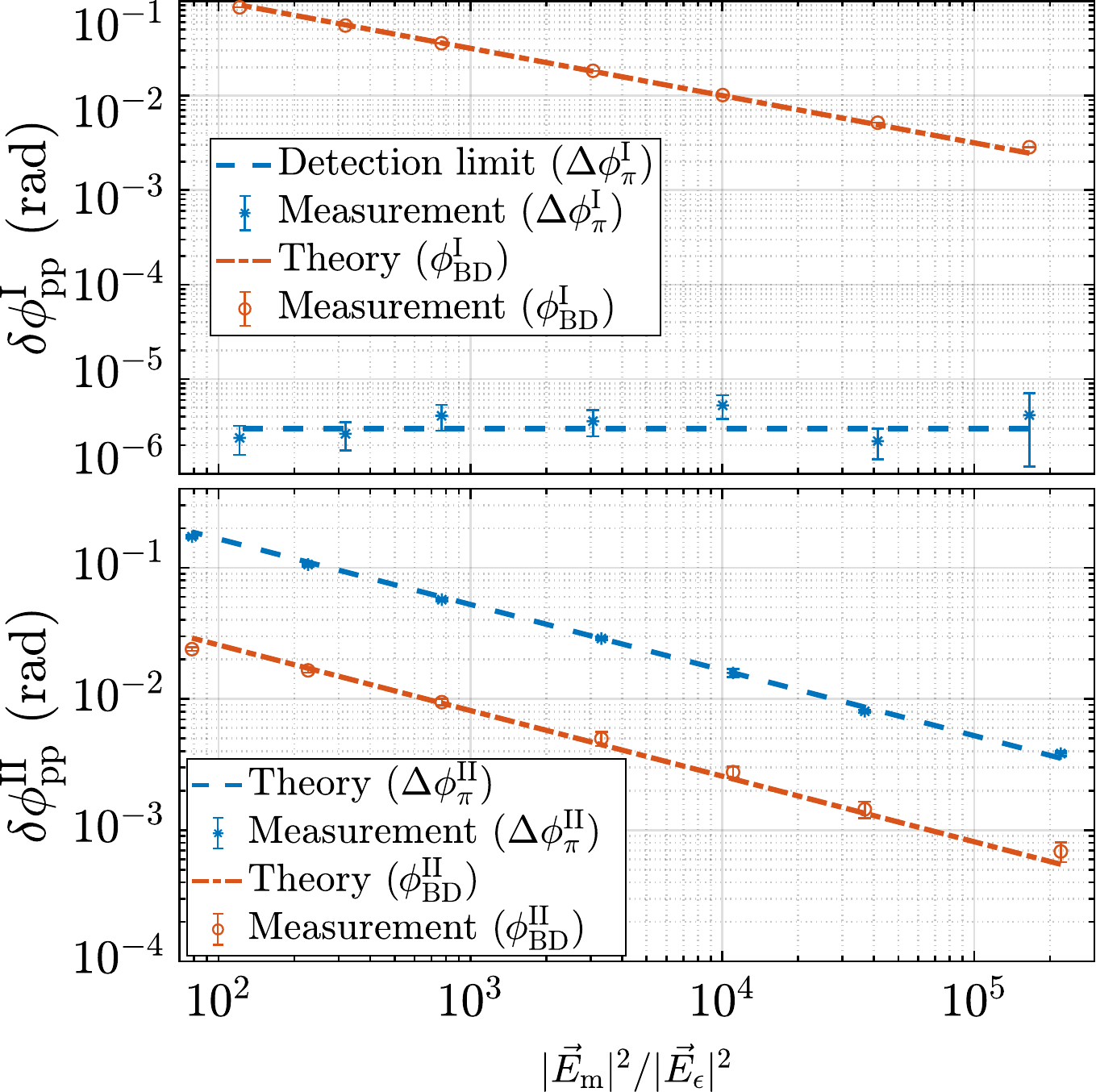}
\caption{Phase noise peak-to-peak amplitude versus relative parasitic beam power in $\pi$ test scheme and BD scheme. Top: type-I parasitic beam. Bottom: type-II parasitic beam. Dashed lines: theoretical prediction. Markers: experimental results. Note that the x-axis is $|\vec{E}_{\rm m}|^{2}/|\vec{E}_{\epsilon}|^2$.}\label{fig: pi_phi_vs_PER}
\end{figure}

Finally, we investigate the third variable: the relative parasitic beam strength with respect to the main beam. We apply the same phase modulation described above, keep the polarization direction fixed at $\theta_{\rm m}=0^{\rm o}$, and vary the parasitic beam power. The resulting peak-to-peak phase as a function of the power ratio between the parasitic and main beams is shown in Fig.~\ref{fig: pi_phi_vs_PER}. As expected, a type-I parasitic beam (top panel) parallel to the main beam does not contribute to the $\pi$ test (blue trace). In all other cases (type-I in balanced detection and type-II in both the $\pi$ test and balanced detection), the phase noise scales with the parasitic field amplitude $|\vec{E}_\epsilon|$, meaning that a two-order-of-magnitude reduction in parasitic beam power reduces the phase noise by one order of magnitude (for a fixed main beam power).

The experimental results agree well with the theoretical models, confirming that both types of parasitic interference degrade performance. Appropriate mitigation strategies can be applied depending on the source. For type-I parasitic beams, birefringence effects can be reduced by tighter alignment of the optical axes of the components in the system, thereby reducing $|\vec{E_{\epsilon}}|$ by limiting polarization leakage. For type-II parasitic beams, improved AR coatings and RF isolation between the AOMs (where applicable) are effective approaches to reduce $|\vec{E_{\epsilon}}|$. However, to further suppress parasitic interference, a systematic mitigation approach based on differential and balanced detection schemes is necessary, as shown in the next section.

\section{Differential heterodyne interferometer \label{S3} }
In this section, we apply the model developed in the previous sections to a differential interferometer~\cite{Zhang:22.QuasiIFO} comprising two interferometers: the target interferometer, which tracks the movable target mirror, and the reference interferometer, which reflects off a fixed mirror. By taking the difference between the two outputs, common-mode noise, such as laser frequency noise and optical path length fluctuations, is largely canceled. From the discussion in Sec.~\ref{S2}, the perpendicular component of the parasitic beam can be mitigated by inserting polarizers. For the parallel component, balanced detection can be used to mitigate type-II parasitic interference. However, type-I parasitic interference has a coupling coefficient of unity (see Fig.~\ref{fig: fBDpi}), i.e., it is not mitigated by balanced detection. In the $\pi$ test case, which is equivalent to the differential interferometer, type-I parasitic interference does not contribute to the phase measurement, but type-II is amplified with a coupling coefficient of approximately two (see Fig.~\ref{fig: f2dp}). Thus, to simultaneously reduce the noise contribution of both types, a differential interferometer with balanced detection and well-polarized inputs offers an effective mitigation strategy.

\begin{figure}[ht]
\centering\includegraphics[width = \linewidth]{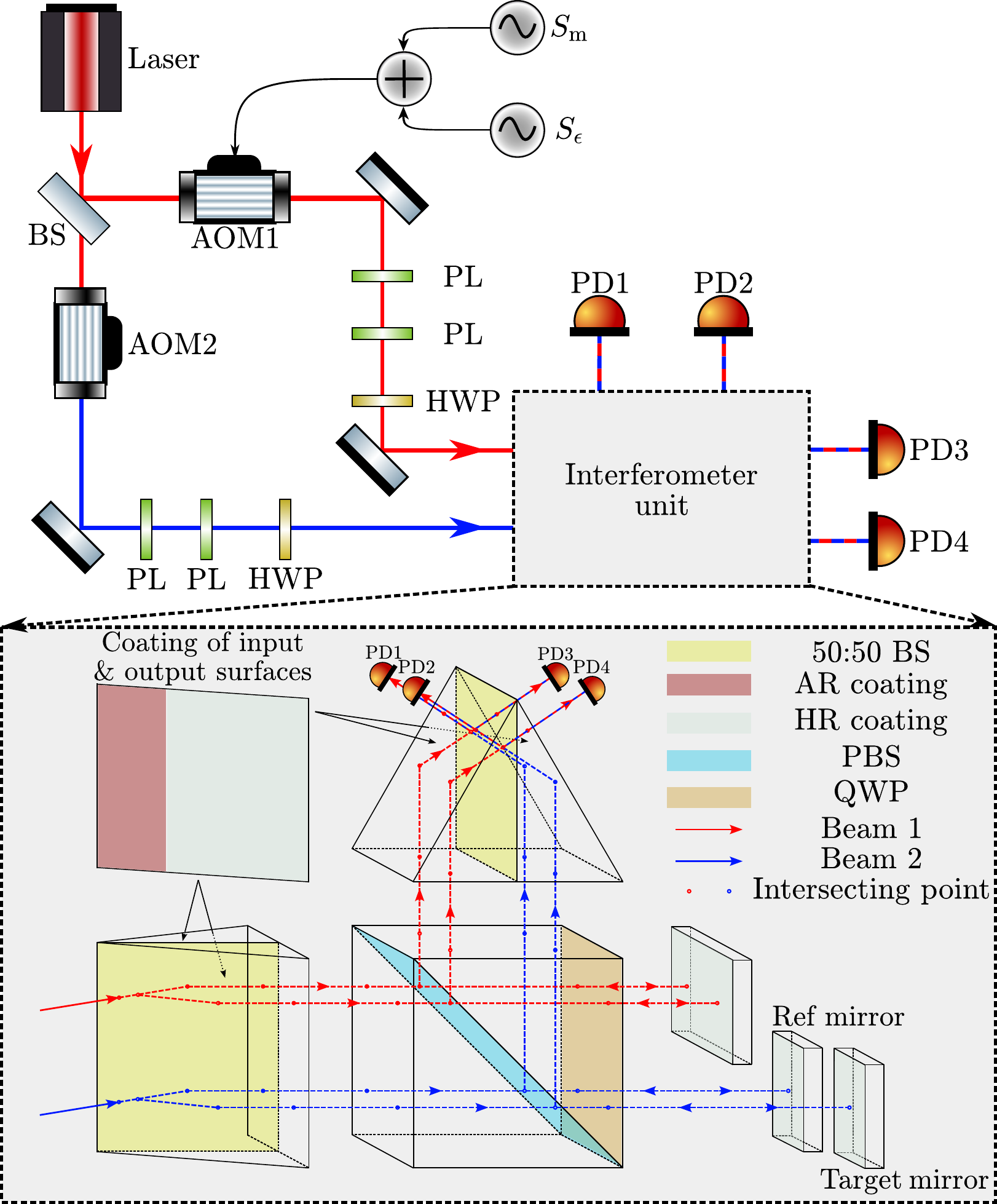}
\caption{Schematic diagram of the setup. The detailed decomposed interferometer unit and the beam paths within are shown at the bottom. PD1 and PD2 detect the relative displacements of the reference and target mirrors with respect to the static mirror, respectively. PD3 and PD4 are implemented for balanced detection. PL: polarizer. HR: high reflection. QWP: quarter-wave plate. PBS: polarizing beam-splitter. HWP: half-wave plate.} \label{fig: ifosetup}
\end{figure}

Figure~\ref{fig: ifosetup} shows the setup used to characterize parasitic interference in a differential interferometer. A 1064\,nm laser output is split and frequency-shifted by two AOMs by 80\,MHz and 81\,MHz, resulting in a heterodyne frequency of 1\,MHz. Two polarizers are placed in each path to provide a polarization extinction ratio (PER) $>10^6$, strongly suppressing the perpendicular polarization component. After the polarizers, half-wave plates are added to control the polarization axes and align them to the polarizing beam splitter inside the interferometer unit. In the interferometer unit (Fig.~\ref{fig: ifosetup} bottom), a differential interferometer is formed comprising the reference and target interferometers, where relative displacements of the reference and target mirrors with respect to the static mirror are detected by the two interferometers, respectively. The former is detected at PD1 and the latter at PD2. The motion of the target mirror relative to the reference mirror is measured from the phase difference between the two interferometers. The additional photodetectors, PD3 and PD4, are required for the balanced detection scheme. For the purpose of this investigation, the target and reference mirrors are implemented using a single mirror, such that in an ideal system, the differential phase is zero, as both interferometers follow identical optical paths. The signals of interest are the single phase $\phi$, differential phase $\Delta\phi$, and differential phase after balanced detection $\Delta\phi_{\rm BD}$, i.e.,
\begin{eqnarray}
  \Delta\phi_{\rm} &=& \phi_{\rm TM} - \phi_{\rm Ref}\label{eq 36}\\
  &=&\angle\vec{V}_2-\angle\vec{V}_1,\ (\mathrm{or}~\angle\vec{V}_4-\angle\vec{V}_3)\nonumber\\
 \Delta\phi_{\rm BD} &=& \phi_{\rm TM}^{\rm BD} - \phi_{\rm Ref}^{\rm BD} \nonumber\\
 &=&[\angle(\vec{V}_2-\vec{V}_4)] - [\angle(\vec{V}_1-\vec{V}_3)],\label{eq 37}\\
 \vec{V}_{k} &=& \exp(i\phi_{k}), \ k=1...4 \label{eq 35}
\end{eqnarray}
where $\phi_{1}$ and $\phi_{3}$ are the phases measured by the reference interferometer at PD1 and PD3, while $\phi_{2}$ and $\phi_{4}$ are those of the target interferometer detected at PD2 and PD4.
\begin{figure}[ht]
\centering\includegraphics[width =\linewidth]{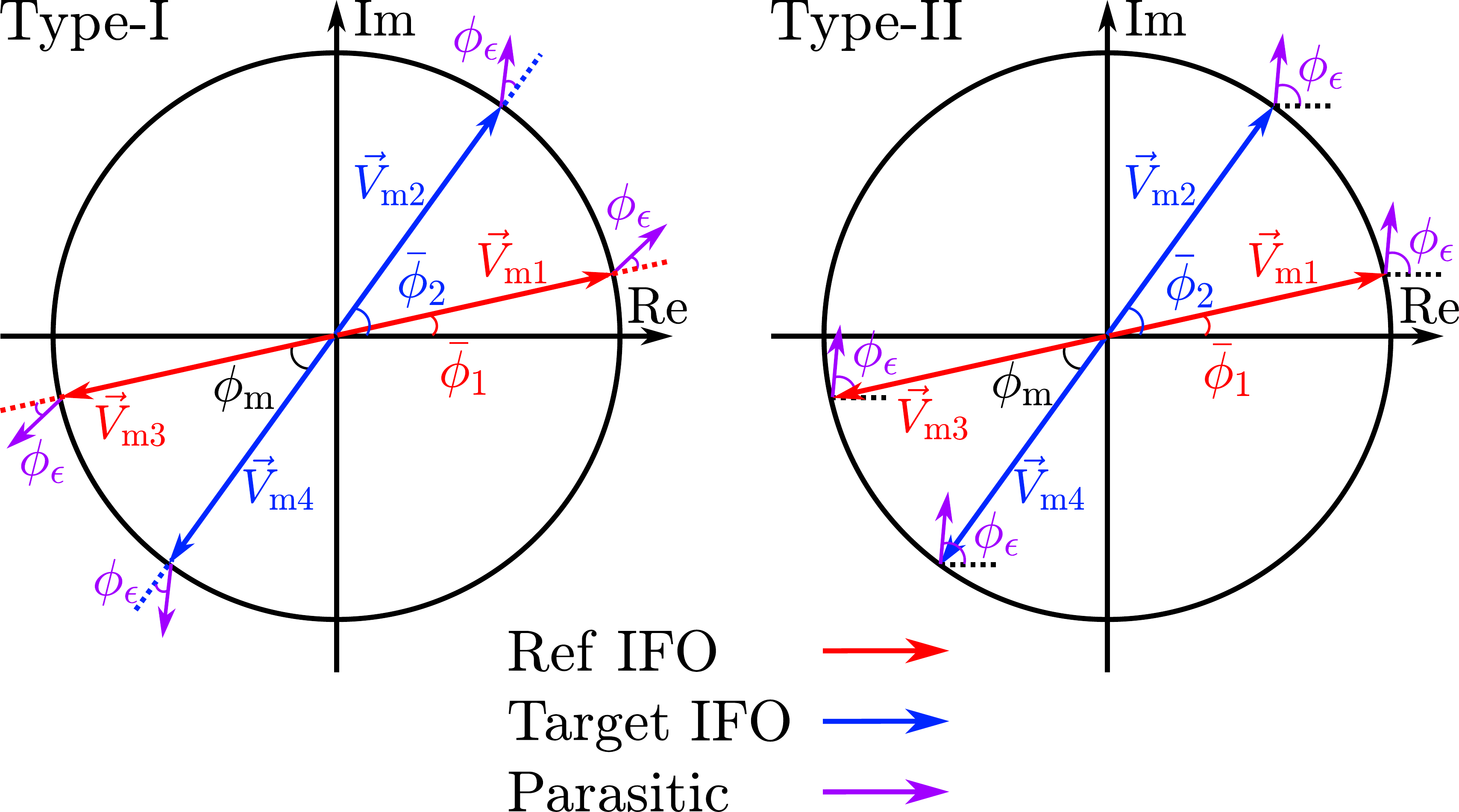}
\caption{Phase vectors acquired by the four PDs with the presence of either type of parasitic interference. Main phase vector pairs with the same color represent the interferometry $\pi$ test. $\phi_{\rm m}$ is the phase difference between the reference and target interferometer without considering the parasitic phase noise. Left: type-I parasitic interference. Right: type-II parasitic interference.} \label{fig: VIFO}
\end{figure}

The phase vectors measured at the four PDs in the presence of type-I parasitic interference are plotted in the complex plane in Fig.~\ref{fig: VIFO} (left). The phase difference between the target and reference interferometers in the absence of parasitic beams is the true measurement $\phi_{\rm m}$ where
\begin{align}
 \phi_{\rm m} = \bar\phi_2-\bar\phi_1=\bar\phi_4-\bar\phi_3.\label{eq 38}
\end{align}
Here, the \ $\bar{}$ \ indicates the true phase without parasitic interference. Note that the contribution of the parasitic phase is independent of $\phi_{\rm m}$, and the resulting noise contribution is
\begin{eqnarray}
 \delta\phi^{\rm I}_{i}&=&f^{\rm I}_{i}(\theta_\epsilon)\frac{|\vec{E}_{\epsilon}|}{|\vec{E}_{\rm m_1}|}\sin\phi_{\epsilon}, \label{eq 39}
\end{eqnarray}
where the subscript $i$ refers to the type of measurement: single, differential, and balanced detection. In contrast, the noise coupling of type-II parasitic interference depends on the true phase $\bar\phi_k$ of each channel and the true measurement $\phi_{\rm m}$, as shown in Fig.~\ref{fig: VIFO} (right). In the previous discussion of type-II parasitic interference in the Mach-Zehnder interferometer (Sec.~\ref{S2}), the noise coupling had a simple form for both differential and balanced detection schemes because $\phi_{\rm m}$ was fixed at $\pi$ by construction. For a real differential interferometer, the noise coupling of type-II is more complex and requires numerical analysis. 
To this end, we developed an analytical model described below.

As shown in Fig.~\ref{fig: ifosetup}, the interferometer unit comprises several optical components. The analytical model propagates the main and parasitic beams through the system separately. The coupling of parasitic interference depends on the polarization and amplitude of the main and parasitic beams at each PD. These are calculated using Jones calculus, where the overall Jones matrix of the system, ${\bf M}^{ij}$, is
\begin{equation}
 {\bf M}^{ij} = {\bf M}_{\rm BS_2}^{ij}\, {\bf M}_{\rm PBS}\,{\bf M}_{\rm QWP}\,{\bf M}_{\rm Mirror}\, {\bf M}_{\rm QWP}\,{\bf M}_{\rm PBS}\, {\bf M}_{\rm BS_1}^{ij}. \label{eq 40}
\end{equation}
where $i$=1,2 and $j$=1,2,3,4. $i$ refers to the two input beams after the AOMs, while $j$ refers to the four PDs. Note that there are eight paths from the two input beams to the four PDs, each described by a different ${\bf M}_{\rm BS_{1,2}}^{ij}$ based on the input and output ports of that path:
\begin{eqnarray}
 {\bf M}^{13}_{\rm BS} &=& \begin{bmatrix} t_{\rm s_1} &0 \\ 0& t_{\rm p_1} \end{bmatrix}; \qquad {\bf M}^{14}_{\rm BS} = \begin{bmatrix} r_{\rm s_1}e^{i\frac{\pi}{2}} &0 \\ 0& r_{\rm p_1}e^{i\frac{\pi}{2}} \end{bmatrix} \nonumber \\
 {\bf M}^{23}_{\rm BS} &=& \begin{bmatrix} t_{\rm s_2} &0 \\ 0& t_{\rm p_2} \end{bmatrix}; \qquad {\bf M}^{24}_{\rm BS} = \begin{bmatrix} r_{\rm s_2}e^{i\frac{\pi}{2}} &0 \\ 0& r_{\rm p_2}e^{i\frac{\pi}{2}} \end{bmatrix}. \nonumber
\end{eqnarray}
where $t$ and $r$ are the transmission and reflection coefficients of field amplitude for s- and p- polarized light. 
Subscripts in $t$ and $r$ indicate coefficients representing beams entering through different ports of the BS. For any input beam with a field amplitude $|\vec{E}_{\rm in}|$ and polarization direction $\hat{\theta}$, the output of the interferometer is
\begin{equation}
 \vec{E}_{\rm out} = {\bf M}^{ij} \vec{E}_{\rm in}\label{eq 41}
\end{equation}
which allows us to simulate the main and parasitic beam fields and their phase vectors (see Fig.~\ref{fig: VIFO}) at the four photodetectors. For the numerical analysis in the following sections, we set $t_{s_{1}}=t_{s_{2}}=\sqrt{0.466}$ and $r_{s_{1}}=r_{s_{2}}=\sqrt{0.530}$, and the polarization of the main beams to $\theta_{\rm m} =0$, nominally aligned to the transmission axis of the PBS. A parasitic beam of either type with a relative power of $10^{-5}$ with respect to the main beam is injected into the system. The field amplitude and phase of the parasitic and the main beams at each PD are simulated using Eq.~(\ref{eq 41}). The phase vectors ($\vec{V}_{\rm{m}_\textit{k}}$, $\vec{V}_{\epsilon_k}$) are then obtained by interfering the fields and extracting the amplitude and phase of the terms at the heterodyne frequency. Applying Eqs.~(\ref{eq 36})---(\ref{eq 35}) with and without the presence of the parasitic phase vector and taking the difference, the phase noise contributions in the three cases of interest are
\begin{eqnarray}
 \delta\phi^{\rm I, II}_{\phi_{k}} &=& \angle\vec{V}^{\rm I, II}_{k}-\angle\vec{V}_{\rm{m}\textit{k}}, \ k=1...4 \label{eq 42}\\
 \delta\phi^{\rm I, II}_{\Delta\phi} &=& \angle\vec{V}^{\rm I, II}_2-\angle\vec{V}^{\rm I, II}_1-(\angle\vec{V}_{\rm m2}-\angle\vec{V}_{\rm m1}) \label{eq 43}\\
 \delta\phi^{\rm I, II}_{\Delta\phi_{\rm BD}} &=& [\angle(\vec{V}^{\rm I, II}_2-\vec{V}^{\rm I, II}_4) - \angle(\vec{V}^{\rm I, II}_1-\vec{V}^{\rm I, II}_3)]\nonumber\\
 &&-[\angle(\vec{V}_{\rm m2}-\vec{V}_{\rm m4}) - \angle(\vec{V}_{\rm m1}-\vec{V}_{\rm m3})], \label{eq 44}
\end{eqnarray}
where 
\begin{align}
    \vec{V}_{\rm{m}\textit{k}}=\exp(i\bar\phi_k),
\end{align}
denotes the phase vectors obtained in the absence of parasitic beams. 
The parasitic phase ($\phi_\epsilon$) and the true phase measurement ($\phi_{\rm m}$) are scanned from $-\pi$ to $\pi$, such that the peak-to-peak phase noise amplitude ($\delta\phi^{\rm I,II}_{\rm pp}$) can be computed. Finally, assuming the sinusoidal dependence on the parasitic phase holds, as confirmed in the following sections, the coupling coefficient is obtained by dividing the relative parasitic field amplitude, i.e.,
\begin{equation}
 f_i^{\rm I, II} = \frac{\delta\phi_{\rm pp, \textit{i}}^{\rm I, II}}{2|\vec{E}_{\epsilon}|/|\vec{E}_{\rm m}|}, \label{eq 47}
\end{equation}
where the factor of two accounts for the peak-to-peak amplitude of a sine or cosine function. The following sections focus on $f^{\rm I,II}_i(\theta)$ for the single channel ($f^{\rm I,II}_{\phi_k}$), differential combination ($f^{\rm I,II}_{\Delta\phi}$), and balanced-detection using all four detectors ($f^{\rm I,II}_{\Delta\phi_{\rm BD}}$). Note that the single-channel case is typically impractical for high-precision measurements due to other noise sources unrelated to parasitic beams.

\subsection{Type-I parasitic interference} \label{S3.1}
\begin{figure}[ht]
\centering\includegraphics[width = \linewidth]{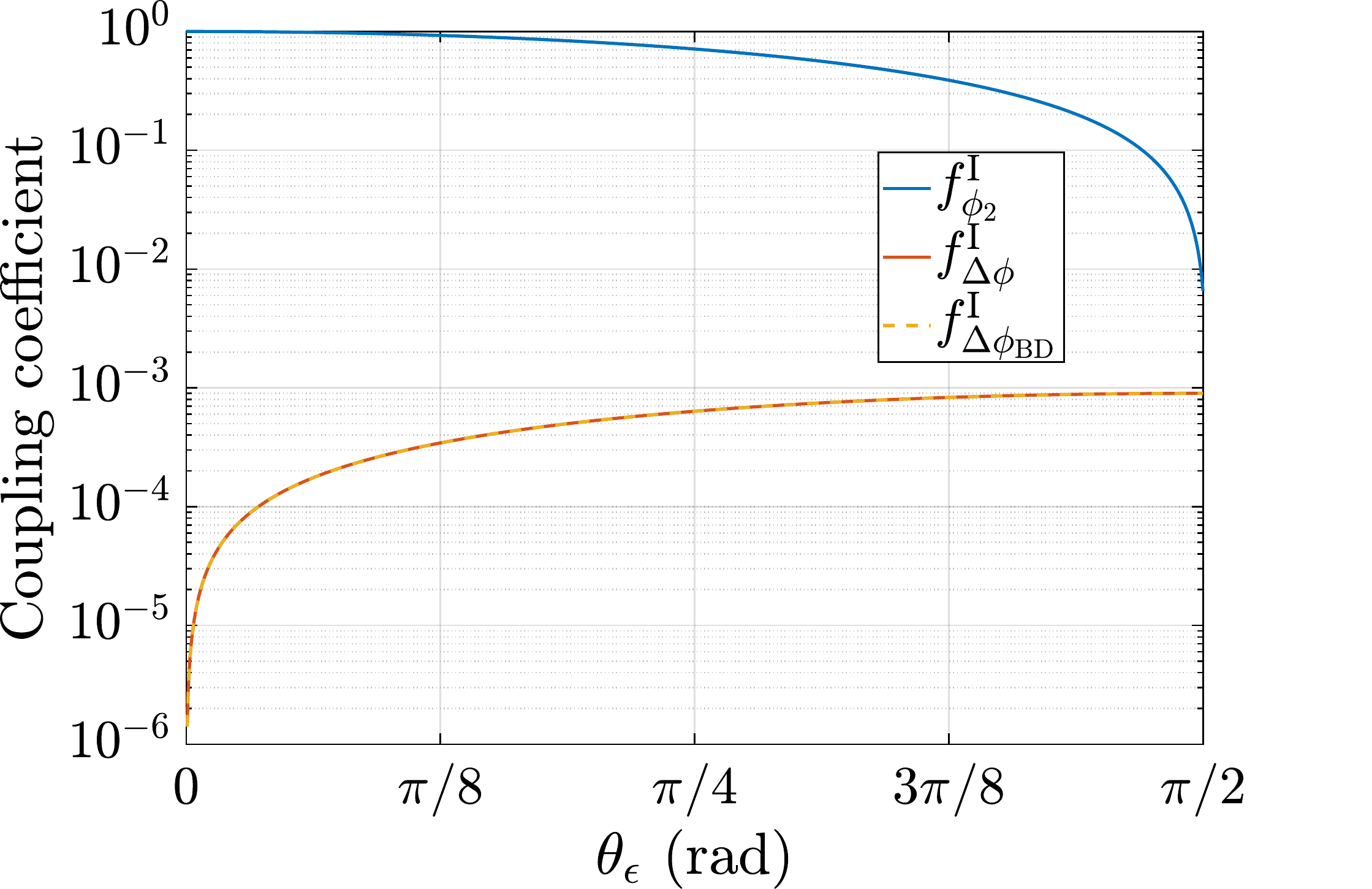}
\caption{Coupling coefficient of type-I parasitic interference versus polarization direction of parasitic beam. $f^{\rm I}_{\phi_2}$: coupling coefficient of a single channel (blue trace). $f^{\rm I}_{\Delta\phi}$: coupling coefficient of the differential phase between two channels (orange trace). $f^{\rm I}_{\Delta\phi_{\rm BD}}$: coupling coefficient of the balanced-detected differential phase using all four channels (dashed yellow trace). The orange trace and the dashed yellow trace coincide.} \label{fig: fI_IFO}
\end{figure}

The simulated coupling coefficient for type-I parasitic interference as a function of the parasitic beam polarization direction is shown in Fig.~\ref{fig: fI_IFO}. For the differential interferometer (red trace), the type-I parasitic interference coupling is greatly reduced compared to the single channel case: the parallel component ($\theta_\epsilon=0$) contributes nothing, while the perpendicular component ($\theta_\epsilon=\pi/2$) has a coupling coefficient of approximately $10^{-3}$. Balanced detection (dashed yellow underneath red trace) offers no improvement over the differential case. The perpendicular component can be attenuated by inserting linear polarizers before the interferometer unit, though rejection depends on their extinction ratio. In summary, differential measurement together with high-quality linear polarizers strongly reduces the effect of type-I parasitic interference. 

\subsection{Type-II parasitic interference} \label{S3.2}
As shown in Fig.~\ref{fig: VIFO} (right), coupling of type-II parasitic interference depends on $\bar\phi_k$. The corresponding phase noise obtained in a single phase measurement is
\begin{equation}\label{eq 46}
 \delta\phi^{\rm II}_{\phi_k}=f^{\rm II}_{\phi_k}(\theta_\epsilon)\frac{|\vec{E}_{\epsilon}|}{|\vec{E}_{\rm m_2}|}\sin(\phi_{\epsilon}-\bar\phi_k). 
\end{equation}
For the differential case, it is simply the noise difference between the single phases, i.e.,
\begin{equation}\label{eq 45}
 \delta\phi^{\rm II}_{\Delta\phi}= f^{\rm II}_{\phi_2}(\theta_\epsilon)\frac{|\vec{E}_{\epsilon}|}{|\vec{E}_{\rm m_2}|}\sin(\phi_{\epsilon}-\bar\phi_2)-f^{\rm II}_{\phi_1}(\theta_\epsilon)\frac{|\vec{E}_{\epsilon}|}{|\vec{E}_{\rm m_2}|}\sin(\phi_{\epsilon}-\bar\phi_1),
\end{equation}
which, assuming $f^{\rm II}_{\phi_1}(\theta_\epsilon)\simeq f^{\rm II}_{\phi_2}(\theta_\epsilon)$, simplifies to 
\begin{equation}\label{eq 52}
 \delta\phi^{\rm II}_{\Delta\phi}\simeq f^{\rm II}_{\Delta\phi}(\theta_\epsilon)\frac{|\vec{E}_{\epsilon}|}{|\vec{E}_{\rm m_2}|}\cos(\phi_{\epsilon}-\bar\phi_1-\phi_{\rm m}/2)\sin(-\phi_{\rm m}/2). 
\end{equation}
It follows that the phase noise of type-II parasitic interference exhibits a sinusoidal dependence not only on the parasitic phase but also on the OPL fluctuation experienced by each beam and the true differential phase $\phi_{\rm m}$. This explains why the so-called cyclic error~\cite{Keem:04} appears as a periodic function of the true phase. Additionally, the coupling coefficient is twice that of the single-channel case ($f^{\rm II}_{\Delta\phi}\simeq 2f^{\rm II}_{\phi_k}$), so the parasitic coupling is amplified in the differential phase measurement. 

To mitigate type-II parasitic interference, the implementation of balanced detection is required. Similar to Eqs.~(\ref{eq.26}) and (\ref{eq.30}), the phase noises obtained in the interferometer pairs (1-3 and 2-4) used for balanced detection are
\begin{eqnarray}
 \delta\phi^{\rm II}_{\rm BD_{13}}&=&f^{\rm II}_{\rm BD_{13}}(\theta_\epsilon)\frac{|\vec{E}_{\epsilon}|}{|\vec{E}_{\rm m_2}|}\sin(\phi_{\epsilon}-\bar\phi_1), \label{eq 50} \\
 \delta\phi^{\rm II}_{\rm BD_{24}}&=&f^{\rm II}_{\rm BD_{24}}(\theta_\epsilon)\frac{|\vec{E}_{\epsilon}|}{|\vec{E}_{\rm m_2}|}\sin(\phi_{\epsilon}-\bar\phi_2), \label{eq 51} 
\end{eqnarray}
with
\begin{eqnarray}
 f^{\rm II}_{\rm BD_{13}}&=& \frac{ f^{\rm II}_{\phi_1}(\theta_\epsilon)- f^{\rm II}_{\phi_3}(\theta_\epsilon)}{2},\\
 f^{\rm II}_{\rm BD_{24}}&=& \frac{ f^{\rm II}_{\phi_2}(\theta_\epsilon)- f^{\rm II}_{\phi_4}(\theta_\epsilon)}{2}.
\end{eqnarray}
As previously, one can assume $f^{\rm II}_{\rm BD_{13}}\simeq f^{\rm II}_{\rm BD_{24}}$ to obtain an analytical expression, i.e.,
\begin{eqnarray}
 &&\delta\phi^{\rm II}_{\Delta\phi_{\rm BD}}= \delta\phi^{\rm II}_{\rm BD_{24}}-\delta\phi^{\rm II}_{\rm BD_{13}} \nonumber \\
 &&\simeq f^{\rm II}_{\rm\Delta\phi_{\rm BD}}(\theta_\epsilon)\frac{|\vec{E}_{\epsilon}|}{|\vec{E}_{\rm m_2}|}\cos(\phi_{\epsilon}-\bar\phi_1-\phi_{\rm m}/2)\sin(-\phi_{\rm m}/2). \nonumber\\
 \label{eq 55} 
\end{eqnarray}
Equations~(\ref{eq 52}) and (\ref{eq 55}) are the approximated analytical solutions for visualizing the dependency of phase noise coupling to different variables. 

\begin{figure}[hb]
\centering\includegraphics[width=\linewidth]{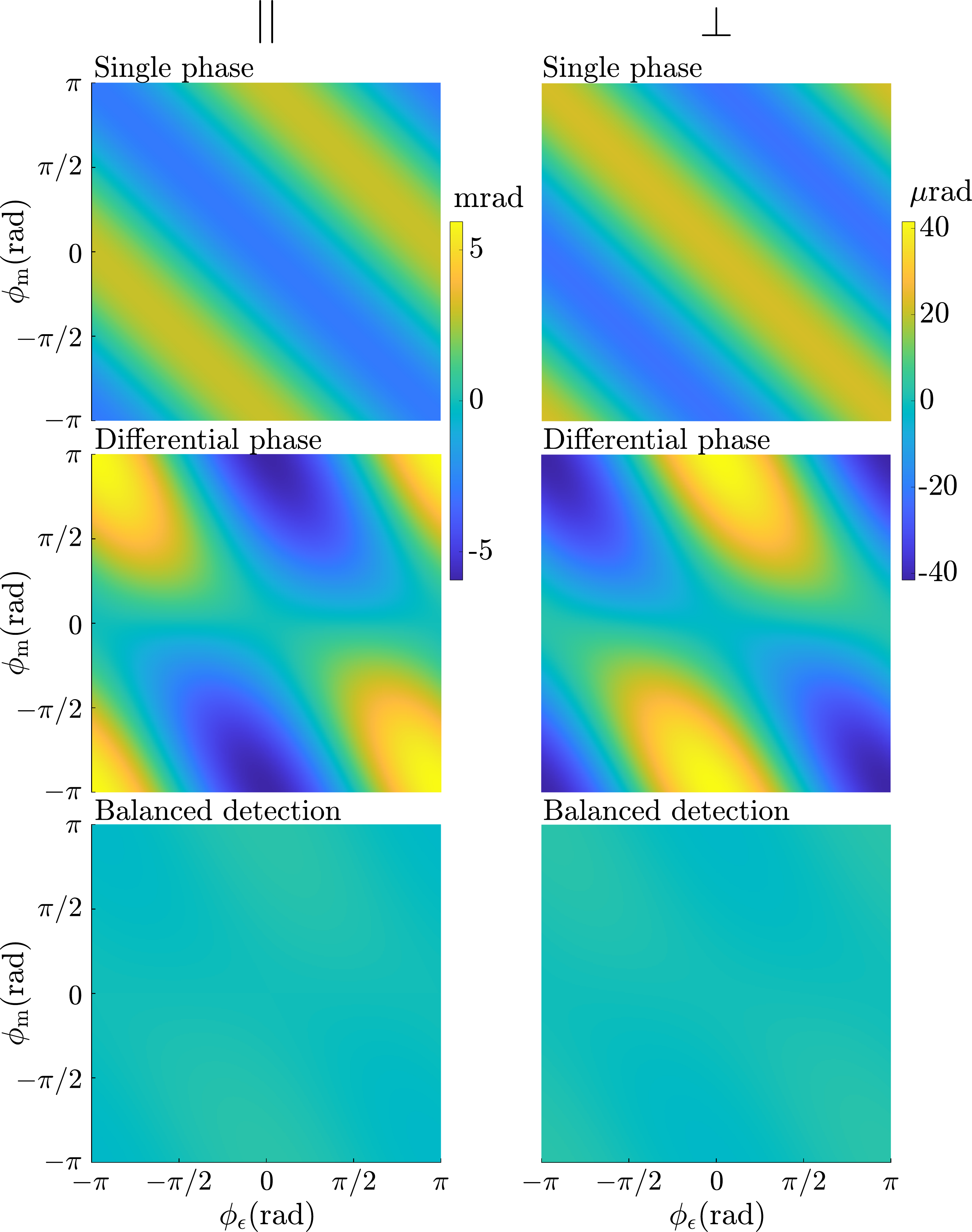}
\caption {Simulation results of phase noise of the parallel (left column) and perpendicular (right column) components of type-II parasitic interference versus parasitic phase ($\phi_{\epsilon}$) and target phase of the measurement ($\phi_{\rm m}$). From top to bottom are the cases of single channel, differential phase, and differential phase with balanced detection, respectively. The results for each polarization component are scaled to the maximum of the differential-phase case.} \label{fig: II_2d}
\end{figure}

Since we assume the parasitic phase is arbitrary, we absorb the fluctuation of $\bar\phi_k$ into the parasitic phase for simplicity ($\bar\phi_k=0$). The simulated phase noise response as a function of $\phi_{\rm m}$ and $\phi_{\rm \epsilon}$ is shown in Fig.~\ref{fig: II_2d} for the single (top), differential (middle), and balanced detection (bottom) schemes, considering both parallel (left) and perpendicular (right) polarization orientations. The response follows a near-sinusoidal dependence on both variables. As $\phi_{\rm m}$ approaches 0\,rad, phase noise with respect to $\phi_{\epsilon}$ greatly reduces, which is expected from Eqs.~(\ref{eq 52}) and (\ref{eq 55}). validating the approximations made earlier. For the parallel case, the single channel, differential, and balanced detection schemes yield maximum parasitic phase amplitudes of 2.97\,mrad, 5.93\,mrad, and 0.41\,mrad, respectively. Compared to the differential phase case, a factor of 15 reduction of the parasitic phase noise is expected with the balanced detection implemented. For the perpendicular case, the effect is a factor of 140 times smaller than the parallel case for the same parasitic beam strength. However, $\rm \mu rad$-level phase noise can be caused by the perpendicular component.

\begin{figure}[ht]
\centering\includegraphics[width = \linewidth]{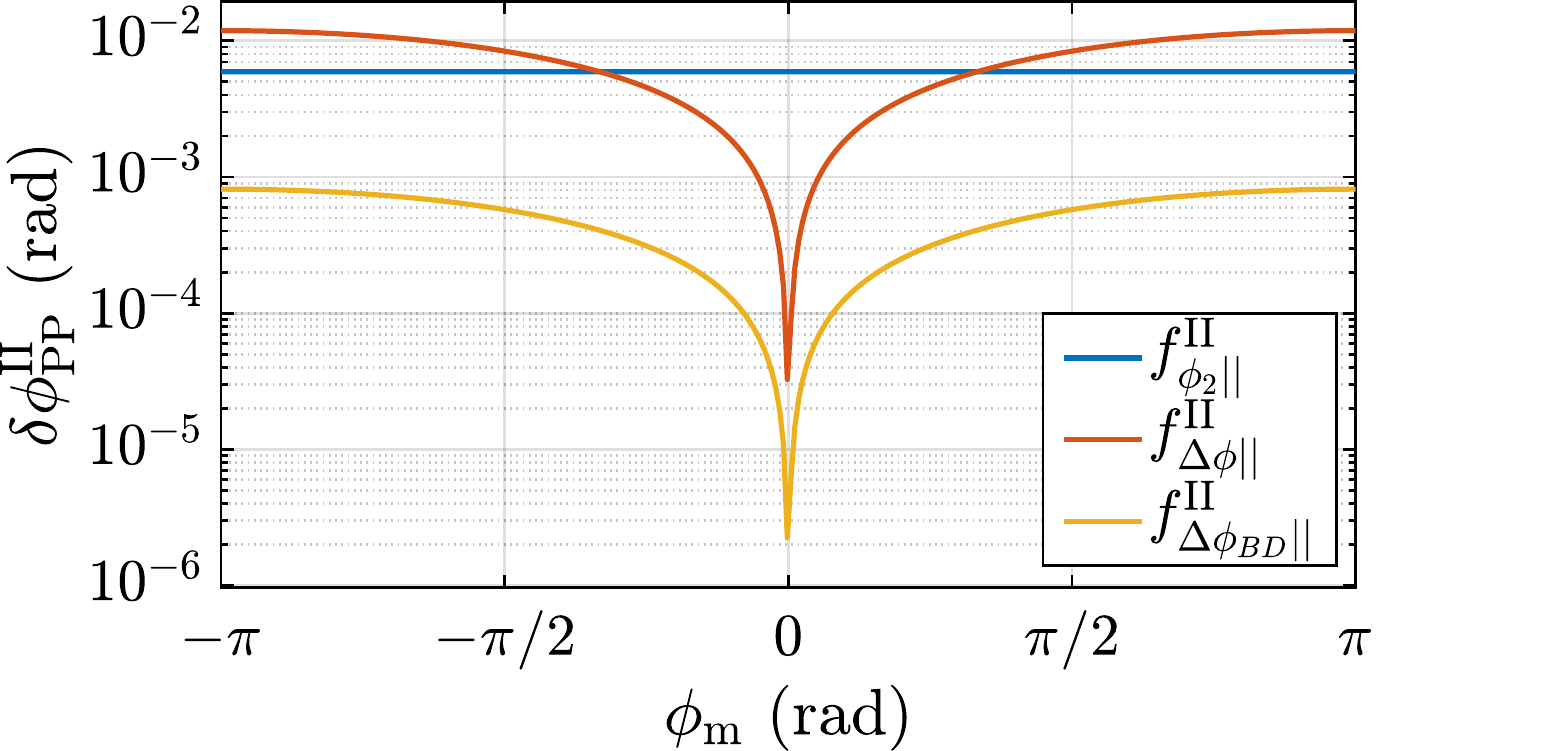}
\caption{Phase noise amplitude of the parallel component of the type-II parasitic interference versus $\phi_{\rm m}$ in three cases: single channel ($f^{\rm II}_{\phi_2||}$), differential phase ($f^{\rm II}_{\Delta\phi||}$), and differential phase with balanced detection ($f^{\rm II}_{\Delta\phi_{\rm BD}||}$).} \label{fig: II_max}
\end{figure}

Since the perpendicular component can be filtered out using polarizers, we focus primarily on the noise contribution of the parallel component. The peak-to-peak phase noise amplitude as a function of $\phi_{\rm m}$ is plotted in Fig.~\ref{fig: II_max}. For the single-phase measurement, $\delta\phi_{\rm pp,\phi_2}$ is independent of $\phi_{\rm m}$. For $|\phi_{\rm m}|\gtrsim 1\,{\rm rad}$, the differential phase measurement amplifies type-II parasitic interference. However, after applying balanced detection, the coupling coefficient is reduced by more than one order of magnitude across all values of $\phi_{\rm m}$. The minimum coupling coefficient occurs at $\phi_{\rm m}=0$ for the differential and balanced detection schemes. Thus, a differential interferometer operating with a small dynamic range, i.e., with $\phi_{\rm m}$ near zero or integer multiples of $2\pi$, significantly mitigates the coupling of type-II parasitic interference. In addition, balanced detection reduces the laser's relative intensity noise (RIN) contribution to phase noise as shown in~\cite{PhysRevApplied.17.024025, PhysRevApplied.20.014016}. For an interferometer with a large dynamic range, the coupling coefficient is proportional to the maximum values shown in Fig.~\ref{fig: II_2d}. Applying Eq.~(\ref{eq 47}), the corresponding coupling coefficients are 0.9, 1.9, and 0.13 for the three cases of interest. Note that the nonlinear nature of the coupling prevents a single coupling coefficient from fully characterizing the noise, as the parasitic phase noise is distributed across multiple frequency bands depending on the spectral content of $\phi_{\rm m}$ and $\phi_\epsilon$. But these coupling coefficients offer estimations of the maximum amount of phase noise that can be induced by type-II parasitic, which allows estimations of the system's susceptibility to type-II parasitic beams. 

\subsection{Beamsplitter} \label{S3.3}
\begin{figure}[hbt]
\centering\includegraphics[width = \linewidth]{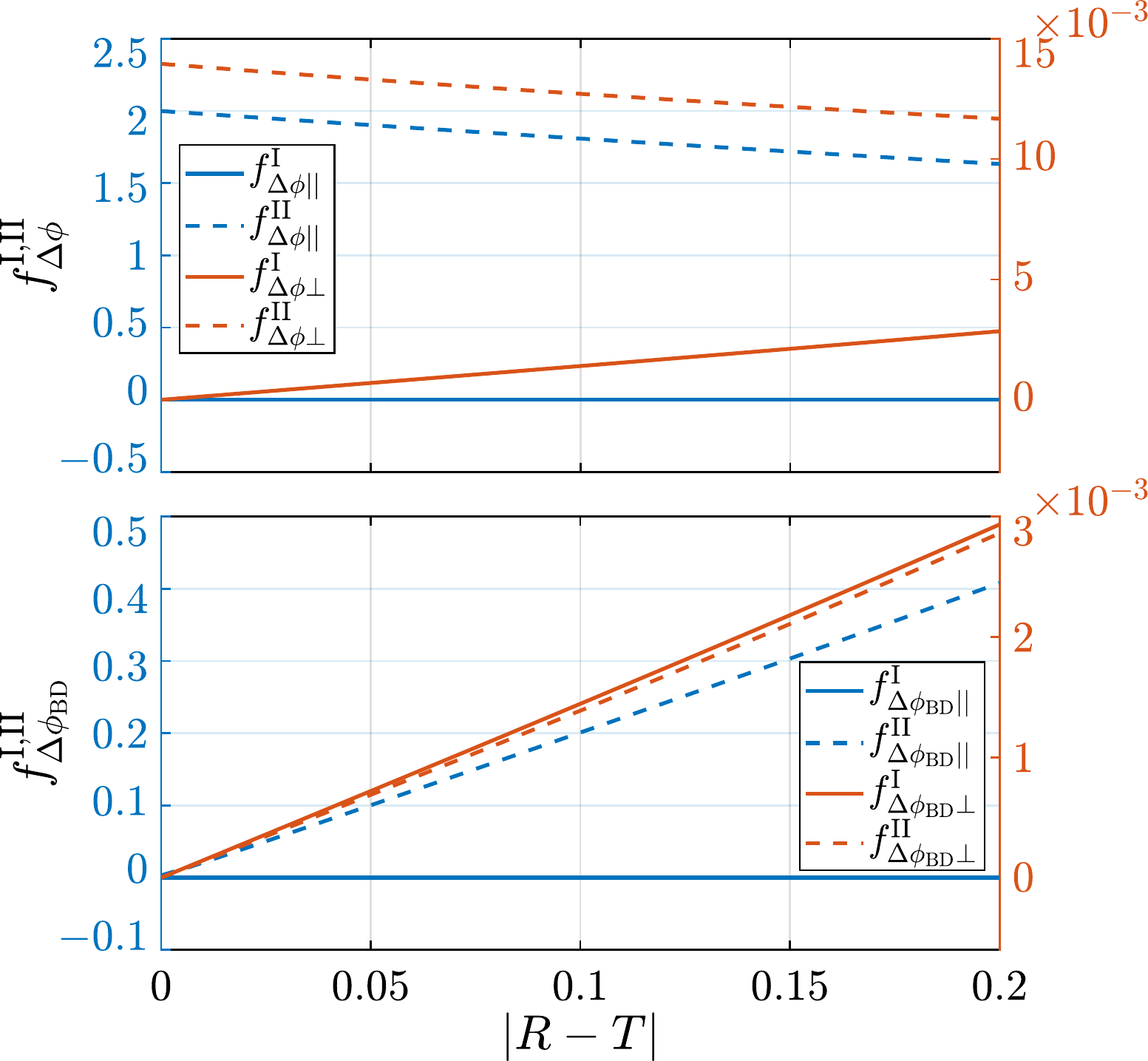}
\caption{Coupling coefficients of the differential phase without (top) and with balanced detection (bottom) applied versus splitting error of the BS. The coupling coefficients of type-II parasitic interference assume the full dynamic range of the interferometer. Left axis: parallel components. Right axis: perpendicular components. Solid line: type-I parasitic interference. Dashed line: type-II parasitic interference. 
\label{fig: f_BS}}
\end{figure}

In addition to the differential and balanced detection schemes, parasitic interference can be further mitigated by other approaches. As discussed in previous sections, the coupling coefficient arises from the polarization dependence of the optical components, particularly in the input and recombination beam splitters. For a perfect 50/50 BS, the transmittance and reflectance are independent of polarization and equal in value. This results in equal relative parasitic beam amplitudes at all channels, such that perfect cancellation is achieved when using a differential interferometer with balanced detection. Figure~\ref{fig: f_BS} shows coupling coefficients (for the parallel and perpendicular components) versus the BS splitting error, defined as the absolute difference between reflectance and transmittance ($|R-T|$). The splitting error is zero for a perfect 50/50 BS and 0.2 for a 60/40 splitting ratio. Here, the coupling coefficients of type-II parasitic interference assume the full dynamic range of the interferometer ($\phi_{\rm m}$ can be any value). In the case of differential phase, as shown in Fig.~\ref{fig: f_BS} (top), the coupling coefficient for type-I parasitic interference vanishes for a perfect BS, whereas type-II parasitic interference persists regardless of the BS quality. With balanced detection (Fig.~\ref{fig: f_BS} bottom), the coupling coefficients scale linearly with the splitting error, and all contributions vanish for an ideal BS. A similar dependence on the beam-splitter splitting ratio has been reported in balanced homodyne detection, where the suppression of classical local-oscillator noise depends critically on the splitting ratio and the balance between the two detection channels~\cite{stefszky2012balanced, chi2015balanced}. This highlights the importance of balanced detection in mitigating parasitic interferences. Note that the coupling coefficients of the perpendicular components also vanish for an ideal BS, indicating that the PER requirement of the input beam can be relaxed with a higher-quality BS. The differential interferometer in our experiment (Fig.~\ref{fig: ifosetup}) has a splitting error of $\sim$0.084.

Polarization alignment of the input beams provides additional means of reducing the coupling of parasitic interference. There are two BSs in the interferometer. The first forms the differential interferometer, and from Fig.~\ref{fig: f1dpperp}, aligning the input polarization to the $\hat{s}$ polarization state with respect to this BS reduces the coupling of type-I parasitic interference. The second is the recombination BS, which determines noise cancellation in balanced detection. Figure~\ref{fig: fBDpi} suggests that aligning the polarization to 45$^{\rm o}$ with respect to the recombination BS greatly reduces the coupling of type-II parasitic interference; in this interferometer, it is designed to be $\hat{p}$-polarized. Adding a half-wave plate before the recombination BS to rotate the polarization by 45$^{\rm o}$ would reduce the type-II coupling coefficient $f^{\rm II}_{\Delta\phi_{\rm BD}}$ by more than a factor of 500.

\subsection{Experimental results} \label{S3.5}
\begin{figure}[ht]
\centering\includegraphics[width = \linewidth]{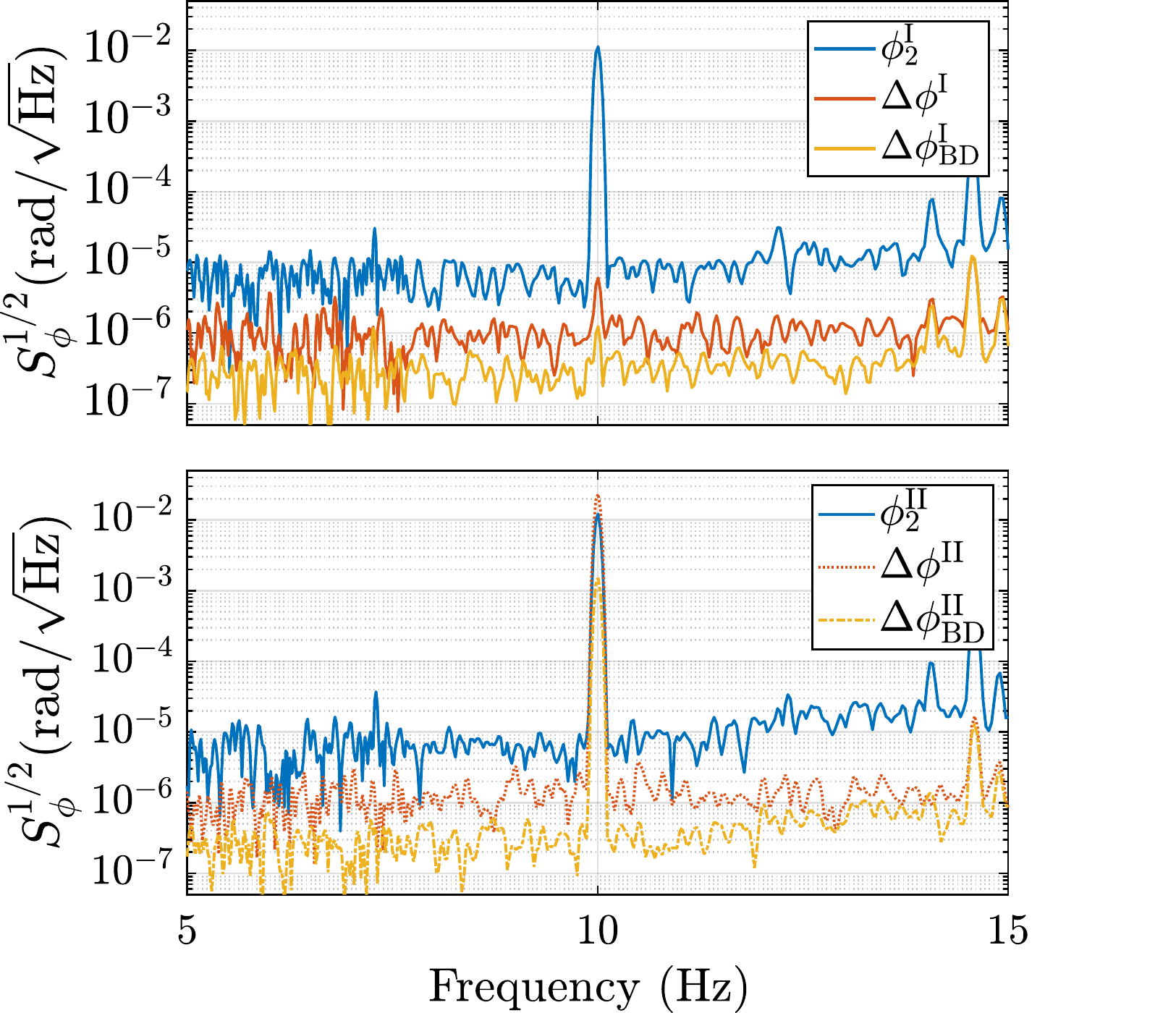}
\caption{Amplitude spectral density plots of measured phase with the presence of a parasitic beam in the differential interferometer system. The parasitic phase is modulated at 10\,Hz, which is responsible for the peaks at 10\,Hz in the spectrum. Top: type-I parasitic interference. Bottom: type-II parasitic interference.} \label{fig: IFO_10Hz}
\end{figure}

Figure~\ref{fig: ifosetup} shows the schematic diagram of the setup. Polarizers are used to provide input beams with PER larger than $10^{6}$, ensuring the perpendicular component of the parasitic beam is well attenuated. Similar to Sec.~\ref{s2.4}, a summing amplifier is used to create an artificial parasitic beam with a relative power of $5\times10^{-4}$ with respect to the main beam. $\phi_\epsilon$ is modulated at 10\,Hz with an amplitude of $\pi/6$ rad, as in the previous experiments. The nominal direction of the parasitic phase vector is tuned to be parallel to the main phase vector in the reference interferometer ($\vec{V}_{\rm m1}$), and $\phi_{\rm m}$ has a nominal value of 2.9\,rad. In this case, the parasitic phase noise appears in the single-channel phase measurement as
\begin{equation}
 \delta\phi_{\phi_{\rm 2}}^{\rm I,II}=f_{\rm \phi_2||}^{\rm I,II}\frac{|\vec{E}_{\epsilon}|}{|\vec{E}_{\rm m}|}\sin\phi_{\epsilon}\simeq \frac{|\vec{E}_{\epsilon}|}{|\vec{E}_{\rm m}|}\sin\phi_{\epsilon},
\end{equation}
where the coupling coefficient is estimated from Figs.~\ref{fig: fI_IFO} and~\ref{fig: II_max}, with a value close to unity for both types of parasitic beams. The sinusoidal term oscillates at 10\,Hz with an amplitude of 0.5. The square root of the power spectral density is shown in Fig.~\ref{fig: IFO_10Hz} for one-minute measurements.

The signals at 10\,Hz in Fig.~\ref{fig: IFO_10Hz} indicate the coupling strength of the parasitic beam. In the single-channel phase ($\phi_2$), the peak is about $11~\rm mrad/\sqrt{Hz}$ for both types of parasitic interference. For the type-I case, simulation from the previous section suggests that type-I parasitic interference is fully removed in a differential interferometer; however, a residual peak of $6~\rm \mu rad/\sqrt{Hz}$ is still present. This is due to RF crosstalk between the AOM drivers when generating type-I parasitic beams, which leads to a residual type-II parasitic beam that leaks into the differential and balanced detection measurements. With balanced detection applied, residual type-II parasitic interference is suppressed, resulting in a peak of $1~\rm \mu rad/\sqrt{Hz}$. Despite this RF crosstalk, the type-I parasitic phase noise is reduced by four orders of magnitude with the differential interferometer (recall that balanced detection does not improve type-I parasitic interference).

For type-II parasitic interference (Fig.~\ref{fig: IFO_10Hz} bottom), phase noise amplitudes of $11\,\rm mrad/\sqrt{Hz}$, $23\,\rm m rad/\sqrt{Hz}$ and $1.5\,\rm m rad/\sqrt{Hz}$ are measured in $\phi$, $\Delta\phi$ and $\Delta\phi_{\rm BD}$, respectively, demonstrating the benefit of balanced detection. Furthermore, the experimental results agree with the theoretical estimates shown in Fig.~\ref{fig: II_max}: a factor of two amplification in the differential phase relative to the single-channel phase, and a factor of 15 improvement with balanced detection.

\begin{figure}[ht]
\centering\includegraphics[width = \linewidth]{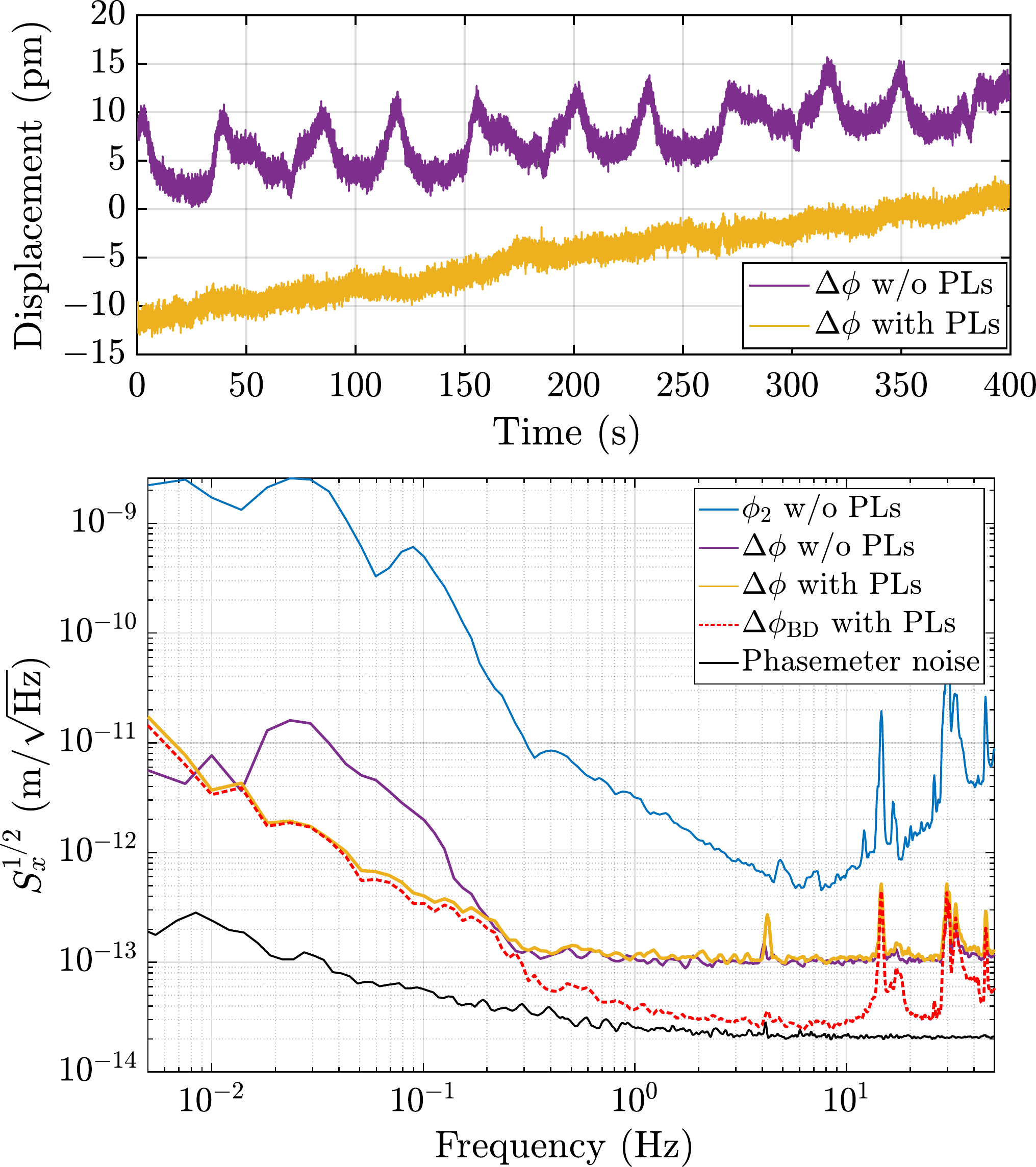}
\caption{Top: time series of the differential phase between reference and target interferometer. Bottom: Amplitude spectral densities of measured displacement for different cases. \label{fig: asd}}
\end{figure}

Finally, we perform measurements without intentionally adding any parasitic beam, using different configurations to mitigate parasitic interference. The whole system, without vacuum isolation, is operated under ambient atmospheric pressure. Thermal shielding is applied to suppress the coupling of temperature fluctuations. Temperature instability below $0.2~\rm mK/\sqrt{Hz}$ is measured for frequencies above 1\,mHz. Additionally, the whole system is covered with an enclosure to suppress airflow disturbances. The phase noise measurement results are shown in Fig.~\ref{fig: asd}. The top panel shows a 400\,s time series segment, where a clear oscillation is present in the absence of polarizers, caused by the perpendicular component of the type-I parasitic beam. The laser source delivers a beam with a PER of 24\,dB, which corresponds to a relative parasitic field amplitude of 0.06. From Fig.~\ref{fig: fI_IFO}, the coupling coefficient of the type-I perpendicular component is $9\times10^{-4}$. Applying Eq.~(\ref{eq 39}) and converting phase to displacement, the estimated peak-to-peak displacement is
\begin{equation}
 \delta \ell_{\rm pp} = \frac{\lambda}{4\pi}\, f^{\rm I}_{\rm \Delta\phi \perp}2\frac{|\vec{E}_{\rm \epsilon}|}{|\vec{E}_{\rm m}|}=9.6\,\rm{pm},
\end{equation}
which agrees well with the experimental data. After implementing linear polarizers, the perpendicular components are strongly suppressed, resulting in a much smoother time series (yellow trace in Fig.~\ref{fig: asd} (top)).

Spectral densities are shown in Fig.~\ref{fig: asd} (bottom). $\phi_2$ (blue trace) shows the fluctuations obtained from a single-channel interferometer. The differential interferometer (purple trace) reduces the coupling of type-I parasitic beams. Linear polarizers (LPs) reduce type-I interference (yellow trace) by about one order of magnitude in the 20\,mHz to 100\,mHz frequency range. For frequencies above 300\,mHz, the system noise floor is limited by type-II parasitic beams. After implementing balanced detection (dashed red trace), displacement noise levels reach 30\,fm/$\sqrt{\rm Hz}$ around a few Hertz, where the system is now limited by the phasemeter noise (black trace). At lower frequencies, residual parasitic interference, alignment jitter, and direct coupling from temperature fluctuations limit the system noise floor.

\begin{figure}[htbp]
\centering\includegraphics[width = .9\linewidth]{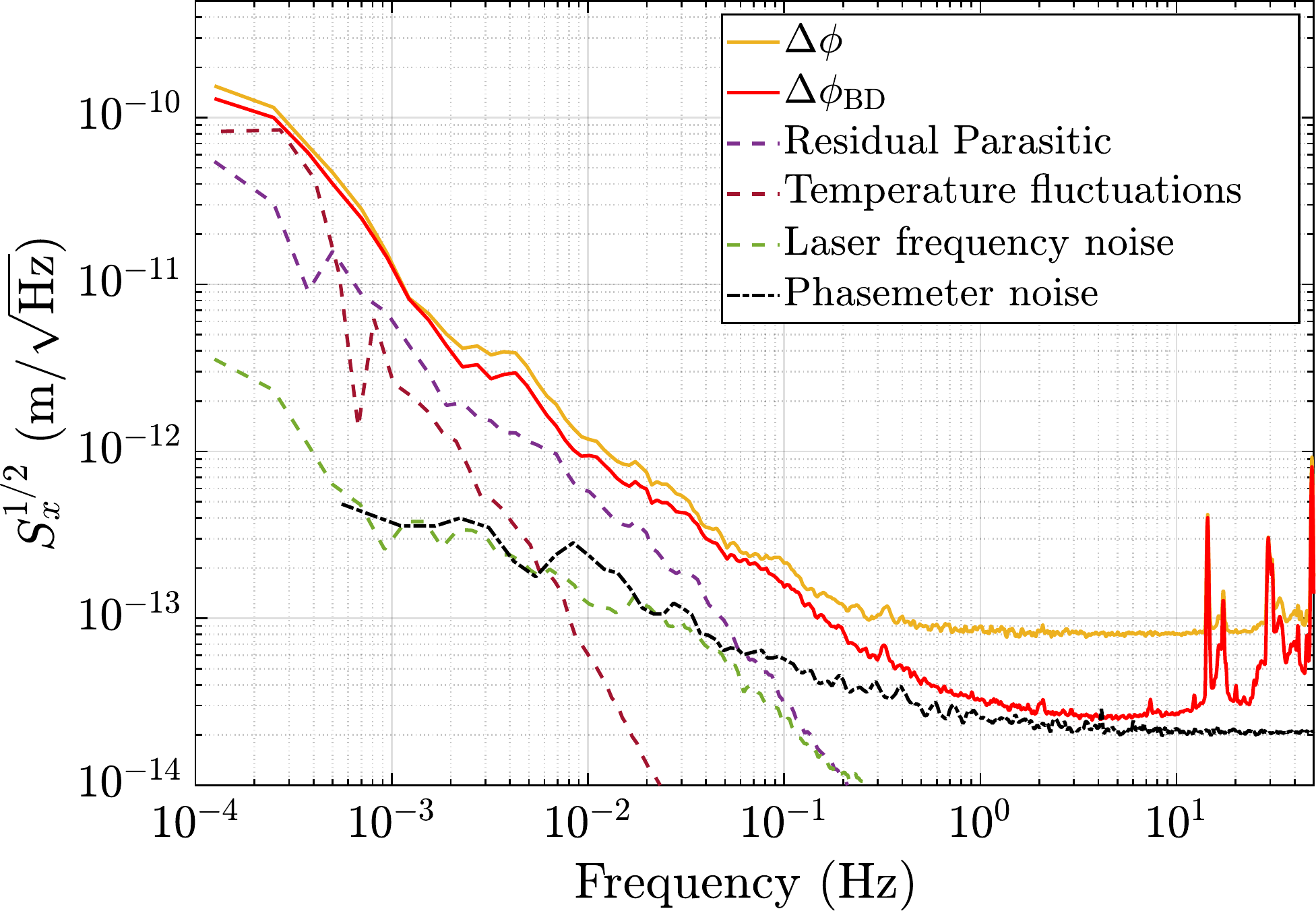}
\caption{Amplitude spectral densities after improvement of input PER and alignment of polarization direction to the AOMs. Noise contributions from different sources are plotted. Measured temperature fluctuations is low-passed at 2\,mHz since it's limited by sensor noise for frequencies above.\label{fig: asdf}}
\end{figure}

To further present the impact of parasitic interference, we reduce the power of the parasitic beams by inserting a polarizer right after the laser source, achieving better alignment of the input polarization to the AOM axis. Compared to the measurements shown in Fig.~\ref{fig: asd}, an order-of-magnitude improvement of the PER in both inputs is achieved, along with less birefringence effect within the AOM crystal. Reduction in the power of both types of parasitic beams is expected. A 2-hour-long measurement is done, and the resulting spectral densities are shown in Fig.~\ref{fig: asdf}.

With a higher PER of the input beams, the noise contribution of the perpendicular components of parasitic beams is further reduced. The differential interferometer (yellow trace) is now limited by type-II parasitic interference, where balanced detection (red trace) improves the noise floor for the frequency band above 1\,mHz. Among all noise sources, residual parasitic interference and phasemeter noise limit the displacement sensitivity for frequencies above 1\,mHz. Within the millihertz to 0.1\,Hz region, residual parasitic interference (dash purple trace) is the main noise source. It is estimated from the beat note amplitude of the single channel where high correlation is observed between the phase measurement and the beat note amplitude. With reduced parasitic beam power, the noise contribution of the residual parasitic interference is further suppressed. Displacement noise floor below 30\,fm/$\sqrt{\rm Hz}$ is achieved for frequencies between 2\,Hz and 10\,Hz, reaching a minimum of 25\,fm/$\sqrt{\rm Hz}$ around 5\,Hz, and it is limited by the phasemeter noise. Below the millihertz frequency regime, the system is limited by the coupling of temperature fluctuations. Nevertheless, suppression of parasitic interference enabled sub-picometer displacement sensitivity at frequencies as low as 9\,mHz.

\section{Design framework for parasitic-interference mitigation}\label{frame work}

\begin{figure*}[bht]
\centering
\begin{tikzpicture}[
    font=\footnotesize,
    mainblock/.style={
        draw,
        rectangle,
        rounded corners=1pt,
        align=center,
        text width=6.0cm,
        minimum height=0.8cm,
        inner sep=2pt
    },
    smallblock/.style={
        draw,
        rectangle,
        rounded corners=1pt,
        align=center,
        text width=2.5cm,
        minimum height=0.7cm,
        inner sep=2pt
    },
    inputblock/.style={
        draw,
        rectangle,
        rounded corners=1pt,
        align=center,
        text width=3.0cm,
        minimum height=0.7cm,
        inner sep=2pt
    },
    decision/.style={
        draw,
        diamond,
        aspect=2.2,
        align=center,
        text width=2.2cm,
        inner sep=1pt
    },
    arrow/.style={
        -{Latex[length=2mm]},
        line width=0.6pt
    }
]
\path[use as bounding box]
(-0.5\textwidth,-9.2cm)
rectangle
(0.5\textwidth,0.5cm);
\begin{scope}[xshift=-6cm]
\node[inputblock] (input) at (1.5,0)
{Define the main fields\\
$(\vec{E}_{\rm m_1},\,\vec{E}_{\rm m_2})$};

\node[mainblock] (block1) at (1.5,-1.3)
{Design the interferometer $(M^{ij})$};

\node[smallblock] (small) at (0,-2.6)
{Estimate parasitic field
amplitudes $|E_\epsilon|$};

\node[mainblock] (block2) at (1.5,-3.9)
{Propagate all fields to the PDs while varying
$\phi_\epsilon$ and $\phi_{\rm m}$};

\node[mainblock] (block3) at (1.5,-5.2)
{Coherently combine the fields with and without the parasitic field, obtain
$\delta\phi_{\rm pp}$};

\node[decision] (decision) at (1.5,-7)
{$\delta\phi_{\rm pp}<\delta\phi_{\rm re}$?};

\node[mainblock] (final) at (1.5,-8.7)
{Final interferometer design};
\end{scope}
\draw[arrow]
(input.south)
--
(block1.north);

\draw[arrow]
([xshift=-1.5cm]block1.south)
--
(small.north);

\draw[arrow]
(small.south)
--
([xshift=-1.5cm]block2.north);

\draw[arrow]
([xshift=1.5cm]block1.south)
--
([xshift=1.5cm]block2.north);

\draw[arrow]
(block2.south)
--
node[midway,right,font=\scriptsize] {$\vec E_{\rm out}$}
(block3.north);

\draw[arrow]
(block3.south)
--
(decision.north);

\draw[arrow]
(decision.south)
--
node[midway,right,font=\scriptsize] {Yes}
(final.north);

\begin{scope}[xshift=-7.5cm]

\node[inner sep=0pt] (table) at (12.0,-4)
{
\renewcommand{\arraystretch}{1.3}

\begin{tabular}{|c|c|c|c|c|}
\hline

&
\multicolumn{2}{c|}{Type-I}
&
\multicolumn{2}{c|}{Type-II}
\\
\hline

&
$\parallel$
&
$\perp$
&
$\perp$
&
$\parallel$
\\
\hline

$|E_\epsilon|$
&
\parbox[c][1.25cm][c]{1.7cm}{
    \centering
    AR coating,\\
    wedge surfaces
}
&
\multicolumn{2}{c|}{
    \parbox[c][1.25cm][c]{3.2cm}{
        \centering
        High PER input
    }
}
&
\parbox[c][1.25cm][c]{1.7cm}{
    \centering
    RF shielding,\\
    wedge surfaces
}
\\
\hline

$f$
&
\multicolumn{2}{c|}{
    \parbox[c][1.25cm][c]{3.2cm}{
        \centering
        Differential interferometer,\\
        $r_p=r_s,\;t_p=t_s$ at BS1
    }
}
&
\multicolumn{2}{c|}{
    \parbox[c][1.25cm][c]{3.2cm}{
        \centering
        Balanced detection,\\
        $r=t$ at BS2
    }
}
\\
\hline

$\phi_\epsilon$
&
\multicolumn{4}{c|}{
    \parbox[c][1.25cm][c]{6.6cm}{
        \centering
        Thermal shielding, vibration isolation,\\
        and polarization alignment
    }
}
\\
\hline

\end{tabular}
};

\end{scope}

\draw[arrow]
(decision.east)
-- node[above,font=\scriptsize] {No}
(table.south |- decision.east)
-- (table.south);

\draw[arrow]
(table.north)
-- ++(0,0.2cm)
|- (block1.east);
\end{tikzpicture}
\caption{
Design and evaluation framework for parasitic interference
in a heterodyne interferometer. $\parallel$: parallel component. $\perp$: perpendicular component. BS1: beam splitter forming the differential interferometer. BS2: recombination beam splitter.
}
\label{fig:design_flowchart}

\end{figure*}

The results presented in the previous sections demonstrate the importance of
accounting for parasitic interference during the design of precision
heterodyne interferometers. Based on the analysis developed in this work, a
systematic framework can be established to evaluate and mitigate parasitic
interference in future interferometer designs.

Figure~\ref{fig:design_flowchart} provides a proposed design procedure
together with the corresponding mitigation strategies. For an interferometer
with a specified phase-sensitivity requirement, an allowable peak-to-peak
parasitic phase error, $\delta\phi_{\rm re}$, can first be defined. The
procedure begins with an initial interferometer configuration and the
definition of the main optical fields. The amplitudes of the relevant
parasitic fields, $|E_\epsilon|$, are then estimated or measured for both
types of parasitic interference and their corresponding polarization
components. For example, the polarization extinction ratio (PER) of the
input beam and the RF crosstalk between the AOMs can be measured to estimate
the amplitudes of the associated parasitic fields.

For a given interferometer configuration, the optical system is modeled by
the corresponding system matrices, $M^{ij}$, incorporating the properties of
the optical components together with realistic alignment errors and
manufacturing tolerances. The main and parasitic fields are propagated to
the photodetectors and coherently combined to obtain the detected heterodyne
signals. By scanning the target phase and the relevant parasitic phases over
$[-\pi,\pi]$, the resulting peak-to-peak parasitic phase error,
$\delta\phi_{\rm pp}$, can be evaluated. Because the parasitic phases are
generally uncontrolled and their coupling to the measured phase is
nonlinear, such a phase scan provides a conservative measure of the
parasitic interference contribution over the considered operating range.

If the predicted contribution does not satisfy the specified requirement,
the design can be iteratively refined by addressing the three quantities
governing parasitic interference: the parasitic field amplitude
$|E_\epsilon|$, the coupling coefficient $f$, and the parasitic phase
$\phi_\epsilon$. The corresponding mitigation strategies are summarized in
Fig.~\ref{fig:design_flowchart}. The parasitic field amplitude can be reduced
through AR coatings, wedged optical surfaces, RF shielding, and polarization
control. In particular, a high input PER suppresses parasitic fields in the
orthogonal polarization component. For a coupling coefficient of order
unity, as is typically the case in the absence of balanced detection or
common-mode rejection, achieving a parasitic phase error of
$1~\mu\mathrm{rad}$ requires a relative parasitic field amplitude on the
order of $10^{-6}$, corresponding to a power ratio on the order of
$10^{-12}$. Suppression based solely on reducing $|E_\epsilon|$ therefore
becomes increasingly challenging as the required phase sensitivity improves,
motivating additional reduction of the coupling coefficient.

As demonstrated in the previous sections, differential interferometry and
balanced detection provide the principal mechanisms for reducing the
coupling coefficients of type-I and type-II parasitic interference,
respectively. For type-I interference, residual coupling is determined by
mismatches between the two nominally identical interferometers. Matching the
polarization-dependent reflection and transmission coefficients of the BS can therefore further improve the common-mode rejection. For
type-II interference, incorporating the measured beat note amplitudes into
the balanced detection scheme can, in principle, provide additional
suppression. In this work, however, amplitude fluctuations
introduced by the photodetectors and ADC limit the accuracy of the amplitude
measurement; therefore, amplitude information is not incorporated in the
balanced detection scheme used in this work and only a factor of 15 suppression of type-II parasitic interference is achieved. Improved amplitude readout,
together with a more balanced splitting ratio at the recombination beam
splitter, could further reduce the type-II contribution.

Finally, fluctuations of $\phi_\epsilon$ can be reduced through thermal
shielding and vibration isolation. Polarization optimization can reduce
birefringence-induced phase variations, while laser frequency stabilization
can reduce parasitic phase fluctuations coupled through the additional
optical path of the parasitic field. These approaches primarily reduce or
shift the parasitic phase fluctuations toward lower frequencies rather than
reducing the intrinsic coupling strength. They are therefore most effective
after $|E_\epsilon|$ and $f$ have first been minimized.

After the relevant mitigation strategies are incorporated, the modified
interferometer is reevaluated to obtain an updated
$\delta\phi_{\rm pp}$. The procedure is repeated until the predicted
parasitic interference contribution satisfies the specified requirement.
The proposed framework therefore allows parasitic interference to be
incorporated as a quantitative design constraint during interferometer
development.

The results presented in the previous sections demonstrate the importance of
accounting for parasitic interference during the design of precision
heterodyne interferometers. Based on the analysis developed in this work, a
systematic framework can be established to evaluate and mitigate parasitic
interference in future interferometer designs.

\section{Summary \label{S4}}
We have derived, simulated, and experimentally validated the effect of parasitic interference in heterodyne interferometry. Parasitic interference is classified into type-I and type-II. In the former, the angular separation between the main and parasitic phase vectors is the same across all channels, while in the latter, parasitic phase vectors in all channels are parallel. These two types encompass all possible ways in which parasitic beams affect the measurement, and their effect on interferometer performance differs. For each type, the parallel and perpendicular polarization components are considered. We have shown that the noise contribution of parasitic beams can be predicted and depends on the coupling coefficient, the relative parasitic beam amplitude, and the phase of the parasitic beam.

We provide systematic mitigation strategies: input beams with high polarization extinction ratio (PER), differential interferometry, and balanced detection are the three key approaches to reducing the coupling of parasitic beams. High-PER inputs eliminate the noise contribution of the perpendicular components. The differential interferometer scheme provides a substantial reduction in type-I parasitic interference. Finally, balanced detection reduces the coupling of type-II parasitic beams.

We validated our models and mitigation strategies first in a simplified Mach-Zehnder interferometer using $\pi$ tests, and second in a differential interferometer with balanced detection implementations. We demonstrate reductions in the coupling coefficients by factors of 10,000 and 10 for type-I and type-II parasitic interference, respectively, and show sub-picometer displacement sensitivity at frequencies as low as 9\,mHz. More importantly, our general theoretical model predicts the experimental results. An evaluation framework for mitigation of parasitic interference in future interferometer design is provided along with detailed mitigation paths, which applies to any heterodyne interferometer system and can be easily extended to homodyne interferometers. 

\section*{Acknowledgments \label{s.ack}}
We thank Dr. Don George and Dr. Jose Rivero for illuminating discussions and support in implementing thermal mitigation measures. We also thank Jackson Dahn for implementing the data acquisition script for temperature measurements. We gratefully acknowledge financial support from the National Aeronautics and Space Administration (NASA) through grants 80NSSC25K7460 and 80NSSC24K1097, and the National Science Foundation (NSF) through grant NSF: PHY-2513439.

\bibliography{apssamp}

@article{Zhang:22.QuasiIFO,
author = {Yanqi Zhang and Felipe Guzman},
journal = {Opt. Lett.},
number = {19},
pages = {5120--5123},
publisher = {Optica Publishing Group},
title = {Quasi-monolithic heterodyne laser interferometer for inertial sensing},
volume = {47},
month = {Oct},
year = {2022},
url = {https://opg.optica.org/ol/abstract.cfm?URI=ol-47-19-5120},
doi = {10.1364/OL.473476},
}

@article{PhysRevApplied.17.024025,
  title = {Relative-Intensity-Noise Coupling in Heterodyne Interferometers},
  author = {Wissel, Lennart and Wittchen, Andreas and Schwarze, Thomas S. and Hewitson, Martin and Heinzel, Gerhard and Halloin, Hubert},
  journal = {Phys. Rev. Appl.},
  volume = {17},
  issue = {2},
  pages = {024025},
  numpages = {18},
  year = {2022},
  month = {Feb},
  publisher = {American Physical Society},
  doi = {10.1103/PhysRevApplied.17.024025},
  url = {https://link.aps.org/doi/10.1103/PhysRevApplied.17.024025}
}

@article{PhysRevApplied.20.014016,
  title = {Influence of Laser Relative-Intensity Noise on the Laser Interferometer Space Antenna},
  author = {Wissel, L. and Hartwig, O. and Bayle, J.B. and Staab, M. and Fitzsimons, E.D. and Hewitson, M. and Heinzel, G.},
  journal = {Phys. Rev. Appl.},
  volume = {20},
  issue = {1},
  pages = {014016},
  numpages = {15},
  year = {2023},
  month = {Jul},
  publisher = {American Physical Society},
  doi = {10.1103/PhysRevApplied.20.014016},
  url = {https://link.aps.org/doi/10.1103/PhysRevApplied.20.014016}
}

@article{WU200317,
title = {Periodic nonlinearity resulting from ghost reflections in heterodyne interferometry},
journal = {Optics Communications},
volume = {215},
number = {1},
pages = {17-23},
year = {2003},
issn = {0030-4018},
doi = {https://doi.org/10.1016/S0030-4018(02)02203-4},
url = {https://www.sciencedirect.com/science/article/pii/S0030401802022034},
author = {Chien-ming Wu},
}

@article{GUO2022110334,
title = {Elimination of parasitic interference effect in fiber-optic external sinusoidal phase-modulating interferometer based on Kalman filter},
journal = {Measurement},
volume = {187},
pages = {110334},
year = {2022},
issn = {0263-2241},
doi = {https://doi.org/10.1016/j.measurement.2021.110334},
url = {https://www.sciencedirect.com/science/article/pii/S0263224121012318},
author = {Jiahui Guo and Xiaojun Liu and Mingliang Hu and Guang Zhou},
}

@article{Liepmann:92,
author = {T. W. Liepmann},
journal = {Appl. Opt.},
number = {28},
pages = {5905--5906},
publisher = {Optica Publishing Group},
title = {Wedged plate beam splitter without ghost reflections},
volume = {31},
month = {Oct},
year = {1992},
url = {https://opg.optica.org/ao/abstract.cfm?URI=ao-31-28-5905},
doi = {10.1364/AO.31.005905},
}

@article{Wu:99,
author = {Chien-ming Wu and John Lawall and Richard D. Deslattes},
journal = {Appl. Opt.},
number = {19},
pages = {4089--4094},
publisher = {Optica Publishing Group},
title = {Heterodyne interferometer with subatomic periodic nonlinearity},
volume = {38},
month = {Jul},
year = {1999},
url = {https://opg.optica.org/ao/abstract.cfm?URI=ao-38-19-4089},
doi = {10.1364/AO.38.004089},
}

@article{Isleif_2017,
doi = {10.1088/1742-6596/840/1/012016},
url = {https://doi.org/10.1088/1742-6596/840/1/012016},
year = {2017},
month = {may},
publisher = {IOP Publishing},
volume = {840},
number = {1},
pages = {012016},
author = {Isleif, K-S and Gerberding, O and Penkert, D and Fitzsimons, E and Ward, H and Robertson, D and Livas, J and Mueller, G and Reiche, J and Heinzel, G and Danzmann, K},
title = {Suppressing ghost beams: Backlink options for LISA},
journal = {Journal of Physics: Conference Series},
}

@article{Isleif_2018,
doi = {10.1088/1361-6382/aaa879},
url = {https://doi.org/10.1088/1361-6382/aaa879},
year = {2018},
month = {mar},
publisher = {IOP Publishing},
volume = {35},
number = {8},
pages = {085009},
author = {Isleif, Katharina-Sophie and Bischof, Lea and Ast, Stefan and Penkert, Daniel and Schwarze, Thomas S and Barranco, Germán Fernández and Zwetz, Max and Veith, Sonja and Hennig, Jan-Simon and Tröbs, Michael and others},
title = {Towards the LISA backlink: experiment design for comparing optical phase reference distribution systems},
journal = {Classical and Quantum Gravity}
}

@article{PhysRevLett.122.081104,
  title = {Picometer-Stable Hexagonal Optical Bench to Verify LISA Phase Extraction Linearity and Precision},
  author = {Schwarze, Thomas S. and Fern\'andez Barranco, Germ\'an and Penkert, Daniel and Kaufer, Marina and Gerberding, Oliver and Heinzel, Gerhard},
  journal = {Phys. Rev. Lett.},
  volume = {122},
  issue = {8},
  pages = {081104},
  numpages = {6},
  year = {2019},
  month = {Feb},
  publisher = {American Physical Society},
  doi = {10.1103/PhysRevLett.122.081104},
  url = {https://link.aps.org/doi/10.1103/PhysRevLett.122.081104}
}

@article{Sasso_2019,
doi = {10.1088/1361-6382/ab0a15},
url = {https://doi.org/10.1088/1361-6382/ab0a15},
year = {2019},
month = {mar},
publisher = {IOP Publishing},
volume = {36},
number = {7},
pages = {075015},
author = {Sasso, C P and Mana, G and Mottini, S},
title = {The LISA interferometer: impact of stray light on the phase of the heterodyne signal},
journal = {Classical and Quantum Gravity},
}

@Article{s18030758,
AUTHOR = {Fu, Haijin and Wang, Yue and Hu, Pengcheng and Tan, Jiubin and Fan, Zhigang},
TITLE = {Nonlinear Errors Resulting from Ghost Reflection and Its Coupling with Optical Mixing in Heterodyne Laser Interferometers},
JOURNAL = {Sensors},
VOLUME = {18},
YEAR = {2018},
NUMBER = {3},
ARTICLE-NUMBER = {758},
URL = {https://www.mdpi.com/1424-8220/18/3/758},
PubMedID = {29498685},
ISSN = {1424-8220},
DOI = {10.3390/s18030758}
}

@Article{s21051708,
AUTHOR = {Gerberding, Oliver and Isleif, Katharina-Sophie},
TITLE = {Ghost Beam Suppression in Deep Frequency Modulation Interferometry for Compact On-Axis Optical Heads},
JOURNAL = {Sensors},
VOLUME = {21},
YEAR = {2021},
NUMBER = {5},
ARTICLE-NUMBER = {1708},
URL = {https://www.mdpi.com/1424-8220/21/5/1708},
PubMedID = {33801264},
ISSN = {1424-8220},
DOI = {10.3390/s21051708}
}

@article{Livas_2017,
doi = {10.1088/1742-6596/840/1/012015},
url = {https://doi.org/10.1088/1742-6596/840/1/012015},
year = {2017},
month = {may},
publisher = {IOP Publishing},
volume = {840},
number = {1},
pages = {012015},
author = {Livas, J and Sankar, S and West, G and Seals, L and Howard, J and Fitzsimons, E},
title = {eLISA Telescope In-field Pointing and Scattered Light Study},
journal = {Journal of Physics: Conference Series},
}

@article{Spector_2012,
doi = {10.1088/0264-9381/29/20/205005},
url = {https://doi.org/10.1088/0264-9381/29/20/205005},
year = {2012},
month = {aug},
publisher = {IOP Publishing},
volume = {29},
number = {20},
pages = {205005},
author = {Spector, Aaron and Mueller, Guido},
title = {Back-reflection from a Cassegrain telescope for space-based interferometric gravitational-wave detectors},
journal = {Classical and Quantum Gravity},
}

@article{Fleddermann_2018,
doi = {10.1088/1361-6382/aaa276},
url = {https://doi.org/10.1088/1361-6382/aaa276},
year = {2018},
month = {feb},
publisher = {IOP Publishing},
volume = {35},
number = {7},
pages = {075007},
author = {Fleddermann, Roland and Diekmann, Christian and Steier, Frank and Tr\"obs, Michael and Heinzel, Gerhard and Danzmann, Karsten},
title = {Sub-pm/rtHz non-reciprocal noise in the LISA backlink fiber},
journal = {Classical and Quantum Gravity},
}

@article{SCHMITZ2009353,
title = {First and second order periodic error measurement for non-constant velocity motions},
journal = {Precision Engineering},
volume = {33},
number = {4},
pages = {353-361},
year = {2009},
issn = {0141-6359},
doi = {https://doi.org/10.1016/j.precisioneng.2008.10.001},
url = {https://www.sciencedirect.com/science/article/pii/S0141635908001219},
author = {Tony L. Schmitz and David C. Chu and Hyo Soo Kim},
}

@article{SCHMITZ2003311,
title = {An investigation of two unexplored periodic error sources in differential-path interferometry},
journal = {Precision Engineering},
volume = {27},
number = {3},
pages = {311-322},
year = {2003},
issn = {0141-6359},
doi = {https://doi.org/10.1016/S0141-6359(03)00036-9},
url = {https://www.sciencedirect.com/science/article/pii/S0141635903000369},
author = {Tony L. Schmitz and John F. Beckwith},
}

@article{abbott2009ligo,
  title={LIGO: the laser interferometer gravitational-wave observatory},
  author={Abbott, Benjamin P and Abbott, R and Adhikari, R and Ajith, P and Allen, Bruce and Allen, G and Amin, RS and Anderson, SB and Anderson, WG and Arain, MA and others},
  journal={Reports on Progress in physics},
  volume={72},
  number={7},
  pages={076901},
  year={2009}
}

@article{acernese2015advanced,
  title={Advanced Virgo: a second-generation interferometric gravitational wave detector},
  author={Acernese, Fausto and Agathos, M and Agatsuma, K and Aisa, Damiano and Allemandou, N and Allocca, Aea and Amarni, J and Astone, Pia and Balestri, G and Ballardin, G and others},
  journal={Classical and Quantum Gravity},
  volume={32},
  number={2},
  pages={024001},
  year={2015},
  publisher={IOP Publishing}
}

@article{Akutsu2019Kagra,
  author = {Akutsu, T. and Ando, M. and Arai, K. and Arai, Y. and Araki, S. and Araya, A. and Aritomi, N. and Asada, H. and Aso, Y. and Atsuta, S. and others},
  title = {KAGRA: 2.5 generation interferometric gravitational wave detector},
  journal = {Nature Astronomy},
  year = {2019},
  volume = {3},
  number = {1},
  pages = {35--40},
  doi = {10.1038/s41550-018-0658-y},
  url = {https://doi.org/10.1038/s41550-018-0658-y},
  isbn = {2397-3366}
}

@article{amaro2017laser,
  title={Laser interferometer space antenna},
  author={Amaro-Seoane, Pau and Audley, Heather and Babak, Stanislav and Baker, John and Barausse, Enrico and Bender, Peter and Berti, Emanuele and Binetruy, Pierre and Born, Michael and Bortoluzzi, Daniele and others},
  journal={arXiv preprint arXiv:1702.00786},
  year={2017}
}

@article{abich2019orbit,
  title={In-orbit performance of the GRACE follow-on laser ranging interferometer},
  author={Abich, Klaus and Abramovici, Alexander and Amparan, Bengie and Baatzsch, Andreas and Okihiro, Brian Bachman and Barr, David C and Bize, Maxime P and Bogan, Christina and Braxmaier, Claus and Burke, Michael J and others},
  journal={Physical review letters},
  volume={123},
  number={3},
  pages={031101},
  year={2019},
  publisher={APS}
}

@Article{mi16010006,
AUTHOR = {Huang, Guangyao and Cui, Can and Lei, Xiaoyang and Li, Qixue and Yan, Shuhua and Li, Xinghui and Wang, Guochao},
TITLE = {A Review of Optical Interferometry for High-Precision Length Measurement},
JOURNAL = {Micromachines},
VOLUME = {16},
YEAR = {2025},
NUMBER = {1},
ARTICLE-NUMBER = {6},
URL = {https://www.mdpi.com/2072-666X/16/1/6},
PubMedID = {39858662},
ISSN = {2072-666X},
DOI = {10.3390/mi16010006}
}

@Article{photonics12121181,
AUTHOR = {Song, Shijun and Liu, Xinyue and Chen, Tao and Liu, Changhua and An, Qichang},
TITLE = {Principles and Applications of Interferometry in Highly Segmented Mirrors Co-Phasing},
JOURNAL = {Photonics},
VOLUME = {12},
YEAR = {2025},
NUMBER = {12},
ARTICLE-NUMBER = {1181},
URL = {https://www.mdpi.com/2304-6732/12/12/1181},
ISSN = {2304-6732},
DOI = {10.3390/photonics12121181}
}

@article{hines2023compact,
  title={Compact optomechanical accelerometers for use in gravitational wave detectors},
  author={Hines, A and Nelson, Andrea and Zhang, Yanqi and Valdes, Guillermo and Sanjuan, Jose and Guzman, Felipe},
  journal={Applied Physics Letters},
  volume={122},
  number={9},
  year={2023},
  publisher={AIP Publishing}
}

@article{hines2020optomechanical,
  title={Optomechanical inertial sensors},
  author={Hines, Adam and Richardson, Logan and Wisniewski, Hayden and Guzman, Felipe},
  journal={Applied optics},
  volume={59},
  number={22},
  pages={G167--G174},
  year={2020},
  publisher={Optical Society of America}
}

@article{NBobroff_1993,
doi = {10.1088/0957-0233/4/9/001},
url = {https://doi.org/10.1088/0957-0233/4/9/001},
year = {1993},
month = {sep},
publisher = {},
volume = {4},
number = {9},
pages = {907},
author = {N Bobroff},
title = {Recent advances in displacement measuring interferometry},
journal = {Measurement Science and Technology},
}

@article{armano2022sensor,
  title={Sensor noise in LISA Pathfinder: An extensive in-flight review of the angular and longitudinal interferometric measurement system},
  author={Armano, Michele and Audley, Heather and Baird, Jonathon and Bin{\'e}truy, Pierre and Born, Michael and Bortoluzzi, Daniele and Brandt, N and Castelli, Eleonora and Cavalleri, Antonella and Cesarini, Andrea and others},
  journal={Physical Review D},
  volume={106},
  number={8},
  pages={082001},
  year={2022},
  publisher={APS}
}

@article{soni2021reducing,
  title={Reducing scattered light in LIGO’s third observing run},
  author={Soni, S and Austin, C and Effler, A and Schofield, RMS and Gonz{\'a}lez, G and Frolov, VV and Driggers, Jennifer C and Pele, A and Urban, AL and Valdes, G and others},
  journal={Classical and Quantum Gravity},
  volume={38},
  number={2},
  pages={025016},
  year={2021},
  publisher={IOP Publishing}
}

@article{soni2024modeling,
  title={Modeling and reduction of high frequency scatter noise at LIGO Livingston},
  author={Soni, Siddharth and Glanzer, Jane and Effler, Anamaria and Frolov, Valera and Gonz{\'a}lez, Gabriela and Pele, Arnaud and Schofield, Robert},
  journal={Classical and Quantum Gravity},
  volume={41},
  number={13},
  pages={135015},
  year={2024},
  publisher={IOP Publishing}
}

@article{PhysRevLett.134.213802,
  title = {Tunable Coherence Laser Interferometry: Demonstrating 40 dB of Stray Light Suppression and Compatibility with Resonant Optical Cavities},
  author = {Voigt, Daniel and Eggers, Leonie and Isleif, Katharina-Sophie and Koehlenbeck, Sina M. and Ast, Melanie and Gerberding, Oliver},
  journal = {Phys. Rev. Lett.},
  volume = {134},
  issue = {21},
  pages = {213802},
  numpages = {9},
  year = {2025},
  month = {May},
  publisher = {American Physical Society},
  doi = {10.1103/PhysRevLett.134.213802},
  url = {https://link.aps.org/doi/10.1103/PhysRevLett.134.213802}
}

@article{DeFreitas01091995,
author = {J.M. De Freitas and M.A. Player},
title = {Polarization Effects in Heterodyne Interferometry},
journal = {Journal of Modern Optics},
volume = {42},
number = {9},
pages = {1875--1899},
year = {1995},
publisher = {Taylor \& Francis},
doi = {10.1080/09500349514551641},
URL = {https://doi.org/10.1080/09500349514551641},
eprint = {https://doi.org/10.1080/09500349514551641}
}

@article{Keem:04,
author = {Taeho Keem and Satoshi Gonda and Ichiko Misumi and Qiangxian Huang and Tomizo Kurosawa},
journal = {Appl. Opt.},
number = {12},
pages = {2443--2448},
publisher = {Optica Publishing Group},
title = {Removing nonlinearity of a homodyne interferometer by adjusting the gains of its quadrature detector systems},
volume = {43},
month = {Apr},
year = {2004},
url = {https://opg.optica.org/ao/abstract.cfm?URI=ao-43-12-2443},
doi = {10.1364/AO.43.002443},
}

@article{Was_2021,
doi = {10.1088/1361-6382/abe759},
url = {https://doi.org/10.1088/1361-6382/abe759},
year = {2021},
month = {mar},
publisher = {IOP Publishing},
volume = {38},
number = {7},
pages = {075020},
author = {Wąs, M and Gouaty, R and Bonnand, R},
title = {End benches scattered light modelling and subtraction in advanced Virgo},
journal = {Classical and Quantum Gravity}
}

@article{Longo_2024,
doi = {10.1088/1361-6382/ad0db0},
url = {https://doi.org/10.1088/1361-6382/ad0db0},
year = {2023},
month = {nov},
publisher = {IOP Publishing},
volume = {41},
number = {1},
pages = {015004},
author = {Longo, Alessandro and Bianchi, Stefano and Valdes, Guillermo and Arnaud, Nicolas and Plastino, Wolfango},
title = {Scattered light monitoring system at the Virgo interferometer: performance improvement and automation based on O3 data},
journal = {Classical and Quantum Gravity}
}

@article{chi2015balanced,
author = {Xiaoli Jin and Jing Su and Yaohui Zheng and Chaoyong Chen and Wenzhe Wang and Kunchi Peng},
journal = {Opt. Express},
number = {18},
pages = {23859--23866},
publisher = {Optica Publishing Group},
title = {Balanced homodyne detection with high common mode rejection ratio based on parameter compensation of two arbitrary photodiodes},
volume = {23},
month = {Sep},
year = {2015},
url = {https://opg.optica.org/oe/abstract.cfm?URI=oe-23-18-23859},
doi = {10.1364/OE.23.023859},
}

@article{stefszky2012balanced,
  author  = {Stefszky, M. S. and Mow-Lowry, C. M. and Chua, S. S. Y.
             and Shaddock, D. A. and Buchler, B. C. and Vahlbruch, H.
             and Khalaidovski, A. and Schnabel, R. and Lam, P. K.
             and McClelland, D. E.},
  title   = {Balanced homodyne detection of optical quantum states
             at audio-band frequencies and below},
  journal = {Classical and Quantum Gravity},
  volume  = {29},
  number  = {14},
  pages   = {145015},
  year    = {2012},
  doi     = {10.1088/0264-9381/29/14/145015}
}
\end{document}